\documentclass[%
 aip,
amsmath,amssymb,
reprint,%
]{revtex4-1}
\usepackage{graphicx}% Include figure files
\usepackage{dcolumn}% Align table columns on decimal point
\usepackage{bm}% bold math
\usepackage{hyperref}
\hypersetup{
    colorlinks = true,
    linkcolor = blue,
    citecolor = blue,
    filecolor = blue, 
    urlcolor = blue, 
   }
\usepackage[utf8]{inputenc}
\usepackage[T1]{fontenc}
\usepackage{mathptmx}
\usepackage{soul}
\usepackage{lineno}
\begin{document}

%\preprint{AIP/123-QED}

\title{Electronic structure and transport in amorphous metal oxide and amorphous metal oxy-nitride semiconductors}
\author{Juhi Srivastava}
%\email[]{juhisri@iitk.ac.in}
\affiliation{Department of Materials Science and Engineering, Indian Institute of Technology Kanpur, Kanpur 208016, UP, India}
\affiliation{Samtel Centre for Display Technologies, Indian Institute of Technology Kanpur, Kanpur 208016, UP, India}
\author{Suhas Nahas}
\affiliation{Department of Materials Science and Engineering, Indian Institute of Technology Kanpur, Kanpur 208016, UP, India}
%\email[]{shsnhs@iitk.ac.in}
\author{Somnath Bhowmick}\email[]{bsomnath@iitk.ac.in}
\affiliation{Department of Materials Science and Engineering, Indian Institute of Technology Kanpur, Kanpur 208016, UP, India}
\author{Anshu Gaur}\email[]{agaur@iitk.ac.in}
\affiliation{Department of Materials Science and Engineering, Indian Institute of Technology Kanpur, Kanpur 208016, UP, India}
\affiliation{Samtel Centre for Display Technologies, Indian Institute of Technology Kanpur, Kanpur 208016, UP, India}
\date{\today}

\begin{abstract}
{
Recently amorphous oxide semiconductors (AOS) have gained commercial interest due to their low-temperature processability, high mobility and areal uniformity for display backplanes and other large area applications. A multi-cation amorphous oxide (\textit{a}-IGZO) has been researched extensively and is now being used in commercial applications. It is proposed in the literature that overlapping In-5\textit{s} orbitals form the conduction path and the carrier mobility is limited due to the presence of multiple cations which create a potential barrier for the electronic transport in \textit{a}-IGZO semiconductors. A multi-anion approach towards amorphous semiconductors has been suggested to overcome this limitation and has been shown to achieve hall mobilities up to an order of magnitude higher compared to multi-cation amorphous semiconductors. In the present work, we compare the electronic structure and electronic transport in a multi-cation amorphous semiconductor, \textit{a}-IGZO and a multi-anion amorphous semiconductor, \textit{a}-ZnON using computational methods. Our results show that in \textit{a}-IGZO, the carrier transport path is through the overlap of outer \textit{s}-orbitals of mixed cations and in \textit{a}-ZnON, the transport path is formed by the overlap of Zn-4\textit{s} orbitals, which is the only type of metal cation present. We also show that for multi-component ionic amorphous semiconductors, electron transport can be explained in terms of orbital overlap integral which can be calculated from structural information and has a direct correlation with the carrier effective mass which is calculated using computationally expensive first principle DFT methods.
}
\end{abstract}

% 
%\pacs{}
%
\maketitle
\section{Introduction}
\begin{figure*}[]
\includegraphics[width=0.75 \linewidth]{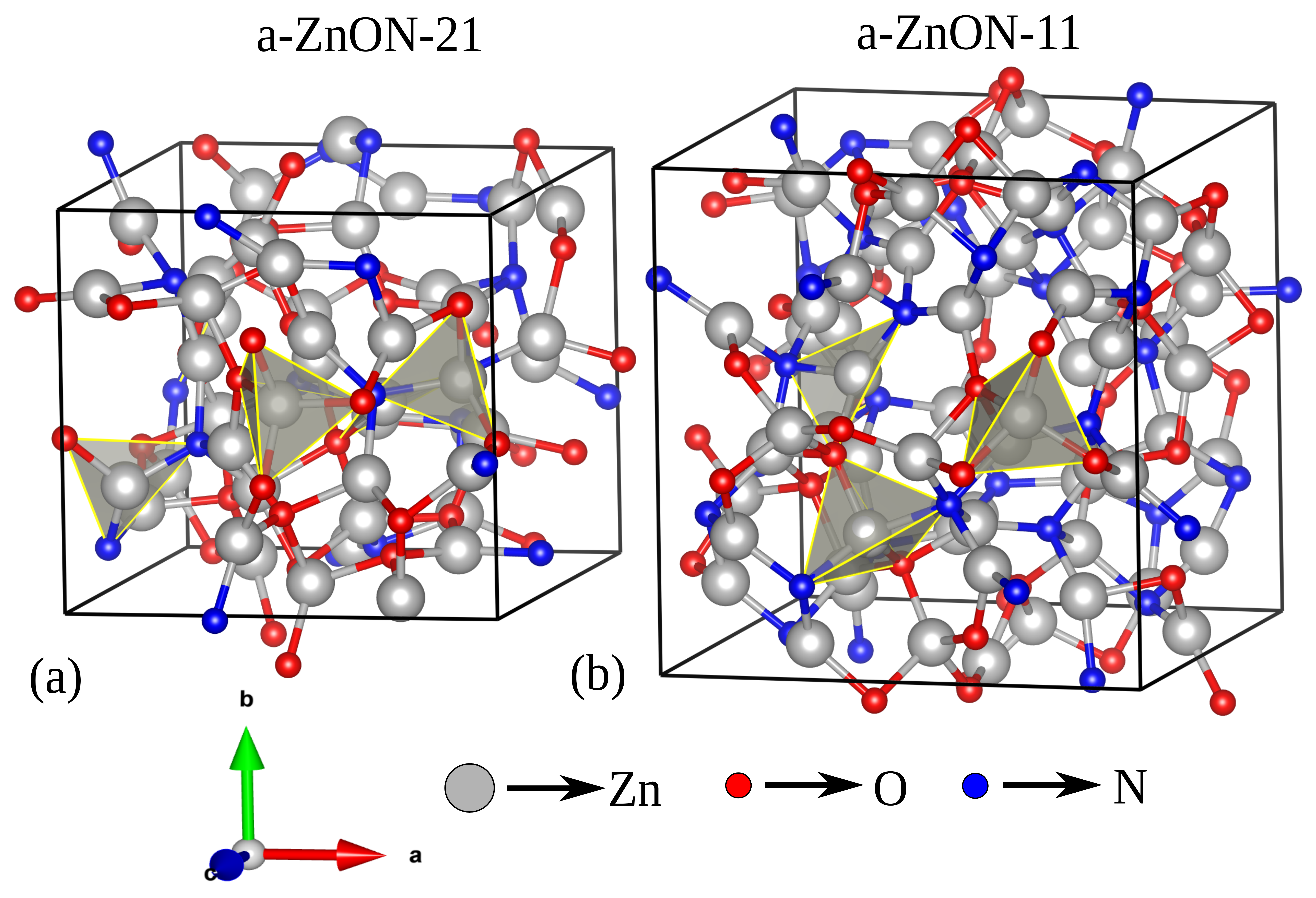}
%\captionsetup{justification=centering}
\caption[]
{Atomic arrangement in (a) \textit{a}-ZnON-21 and (b) \textit{a}-ZnON-11. The polyhedra shown in unit cells represent different coordination of Zn with the anions (O/N).  Colors used for atoms are as follows: Zinc- grey, Oxygen- red and Nitrogen-blue. Visualization software VESTA \cite{46} is used for the visualization of the amorphous structures and polyhedra formations.}	
\label{fig.1}
\end{figure*}

Large area thin film device applications such as active-matrix (AM) flat panel displays and flexible electronics require high uniformity with low processing temperatures.
Hydrogenated amorphous silicon (\textit{a}-Si:H) had been used widely as the active channel material in thin film transistors (TFTs) because of its low fabrication cost and high areal homogeneity up until the last decade \cite{1, 2, 3, 4}. Since 2010, \textit{a}-Si:H based thin film devices have been actively replaced with either low-temperature polysilicon (LTPS) which offers higher mobility but suffers from non-homogeneity \cite{56, 57} or amorphous metal-oxide based TFTs for large area applications such as display backplanes. Over the past decade, metal oxide semiconductors have been studied intensively and used in many thin-film based devices \cite{5, 6, 7, 8, 9, 10}. In case of polycrystalline single metal oxide semiconductors, such as ZnO, SnO$_2$, In$_2$O$_3$, Ga$_2$O$_3$ etc., the conduction band is made of large overlapping \textit{s}-orbitals of metal cations (with electronic configuration (n-1)\textit{d}$^{10}$n\textit{s}$^0$). The overlap between metal \textit{s}-orbitals remain insensitive to any M-O-M bond angle variation on amorphization. Following the hypothesis proposed by Hosono \textit{et al.}\cite{11, 12}, amorphous structures of multi-cation metal oxides are preferred for large areal uniformity. Due to non-directional nature of ionic bonding in these materials, the charge carrier mobility is preserved on amorphization, in comparison to conventionally used \textit{a}-Si, where the directional nature of covalent bonds deteriorates the mobility in amorphous structures\cite{13}. One such multi-cation amorphous oxide semiconductor is amorphous indium gallium zinc oxide (\textit{a}-IGZO), which has shown Hall mobilities as high as 15 cm$^2$/V.sec \cite{13, 14, 15, 16, 17} and is now being adopted by the display industries. \textit{a}-IGZO based TFTs have found application in various areas, such as large area AMOLED displays \cite{18, 19}, high-resolution photo-sensors \cite{20, 21}, flexible TFTs \cite{22, 23}, interactive displays \cite{24, 25} etc. However, it is claimed that due to the presence of multiple cations of different ionic sizes, the electronic conduction path is hindered, which limits the carrier mobility \cite{26}. As an alternative to multi-cation amorphous metal oxides, a multi-anion approach towards amorphous semiconductors has been proposed and is being investigated experimentally and computationally \cite{26, 27, 28, 29, 30, 31, 32, 33}. Amorphous zinc oxy-nitride (\textit{a}-ZnON), as a multi-anion amorphous semiconductor, has shown a great promise as a viable replacement of \textit{a}-IGZO and the electron mobilities (Hall mobilities) exceeding 200 cm$^2$/V.s have been experimentally reported \cite{33}. Challenges related to stability of the \textit{a}-ZnON compositions and fabrication of thin film transistors utilizing \textit{a}-ZnON as active layer are being addressed experimentally and continuous improvements in device structures and device characteristics have been reported \cite{30, 33}.
To realize the full potential of single cation multi-anion approach towards amorphous semiconductors, it is essential to understand theoretically how the electronic conduction is differentiated between multi-cation and multi-anion amorphous semiconductors. It is also necessary to comprehend how to approach complex multi-component amorphous structures computationally, to optimize properties that are important to fabricate stable, high-mobility thin-film transistors. 
In the present work, we have studied computationally generated structures of multi-anion amorphous material \textit{a}-ZnON and compared their electronic structure with that of multi-cation \textit{a}-IGZO. The electronic structures of the two materials are calculated using first-principle density functional theory (DFT), to understand key differences between the two and to understand the relation between structural and electronic properties of multicomponent amorphous materials.
The paper is organized as follows; In Sec-II, the structural details of computationally generated \textit{a}-ZnON structures are discussed, emphasizing that the generated structures of different compositions are indeed amorphous in nature. Following this, in Sec-III, the electronic structure of \textit{a}-ZnON is examined and compared with that of multi-cation \textit{a}-IGZO to highlight the differences in electronic conduction in multi-cation and multi-anion amorphous semiconductors. The interdependence of structural and electronic properties of multi-component amorphous materials is described in Sec-IV, in which direct correspondence of the structural details with physically measurable quantities like electron mobility and the effective mass is accentuated. Conclusions drawn based on this work are summarized in Sec-V.

\begin{figure*}[]
\includegraphics[width=1.0 \linewidth]{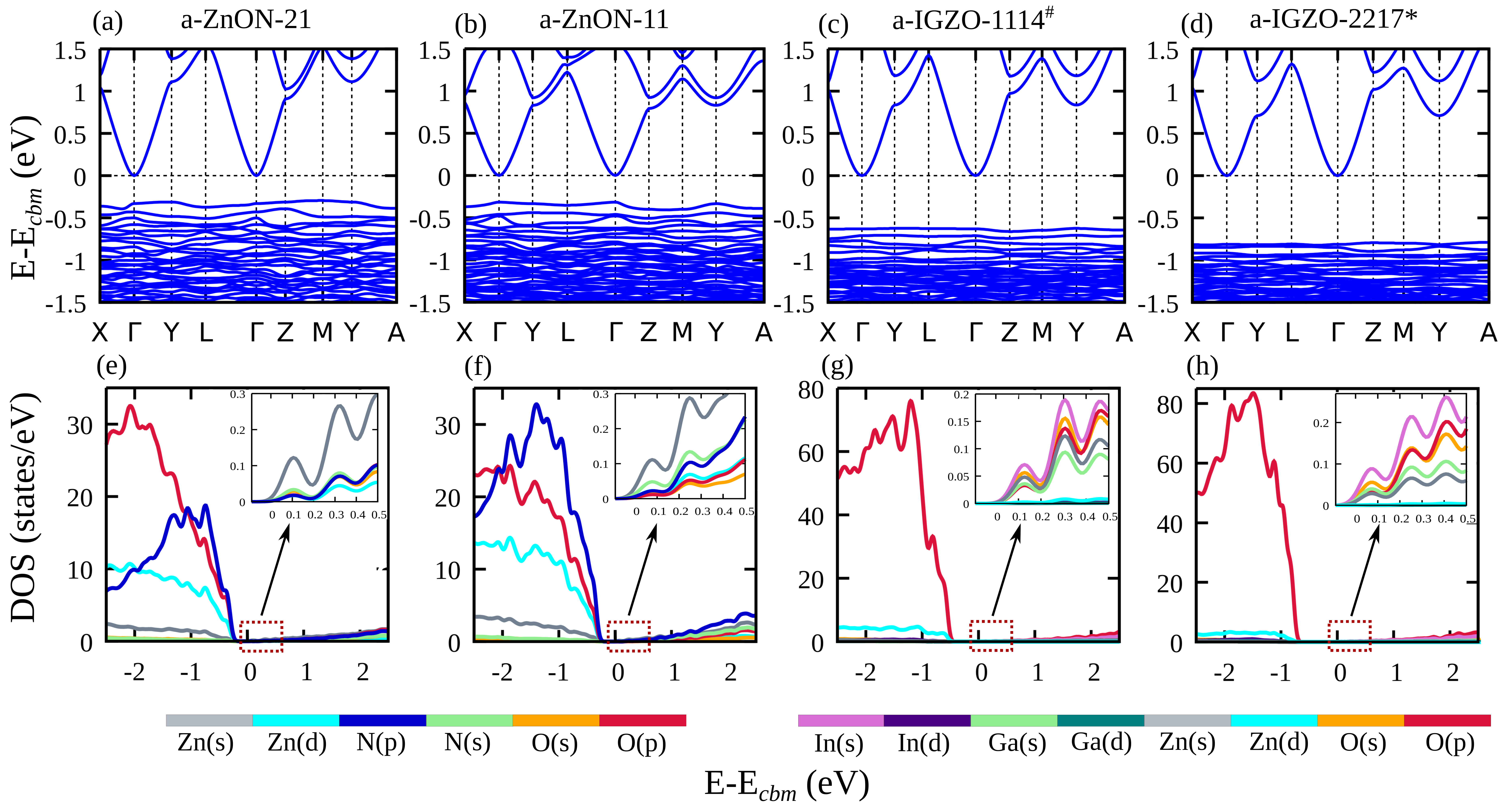}
%\captionsetup{justification=centering}
\caption[]
{(Color online) Electronic pseudo-band structure and PDOS plots for (a) \textit{a}-ZnON-21 and (b) \textit{a}-ZnON-11 compositions and for (c) \textit{a}-IGZO-1114 and (d) \textit{a}-IGZO-2217 compositions. These plots are constructed for a representative structure from each type and composition. Colors used for different orbitals are shown below using a qualitative color bar.}	
\label{fig.2}
\end{figure*}
    
\section{Structure of amorphous zinc oxy-nitride (\lowercase{\textit{a}}-Z\lowercase{n}ON)}
%\begin{figure*}[]
%\includegraphics[width=0.75 \linewidth]{./figures/ZnON_3poly_2_white.pdf}
%%\captionsetup{justification=centering}
%\caption[]
%{Atomic arrangement in (a) \textit{a}-ZnON-21 and (b) \textit{a}-ZnON-11. The polyhedra shown in unit cells represent different coordination of Zn with the anions (O/N).  Colors used for atoms are as follows: Zinc- grey, Nitrogen- red and Oxygen-red. Visualization software VESTA \cite{46} is used for the visualization of the amorphous structures and polyhedra formations.}	
%\label{fig.1}
%\end{figure*}

Amorphous structures of zinc oxy-nitrides (Zn$_{x+1.5y}$O$_x$N$_y$) were generated for two stoichiometries  of x:y (O:N) ratios of 2:1 (Zn$_{42}$O$_{24}$N$_{12}$) and 1:1 (Zn$_{60}$O$_{24}$N$_{24}$), hereafter referred to as \textit{a}-ZnON-21 and \textit{a}-ZnON-11, respectively. We used an evolutionary algorithm based code USPEX (Universal Structure Predictor: Evolutionary Xtallography)\cite{34, 35, 36} to generate the amorphous structures following the methodology proposed by Nahas \textit{et al.}\cite{37}. A large number of structures, obtained from USPEX for each composition, were sorted according to their energies and the structures having no undesirable bonds and densities $>$ 85\% of the theoretical density of corresponding crystalline structures were selected for further analysis. Details of the structure generation and sorting are provided in the  supplementary information (SI-I) and the representative unit cells for both the compositions of \textit{a}-ZnON are shown in Fig.\ref{fig.1}.
  
  We performed a detailed structural analysis of selected structures to calculate average bond lengths (ABL) and average coordination number (ACN). The average bond lengths for Zn-O and Zn-N bonds in both the compositions of \textit{a}-ZnON were found to be very close to the bond lengths in \textit{c}-ZnO and \textit{c}-Zn$_3$N$_2$. The polyhedra shown in Fig.\ref{fig.1} suggest that most of the Zn atoms have either 3 or 4 coordination with anions (O/N). The average coordination number (ACN) of Zn with anions was found to be $\sim$ 3.5 for both compositions, which is slightly less than the values of \textit{c}-ZnO and \textit{c}-Zn$_3$N$_2$ (both having 4 coordinated Zn atoms with respective anions \cite{42, 44, 47}). We attribute this difference to the fact that amorphous structures have densities in the range of 85\%-91\% of the theoretical density, resulting in more open structures. We also calculated radial distribution function (RDF) and running coordination number (RCN) for the selected amorphous structures and found absence of any long-range order and hence these structures were confirmed to be amorphous in nature. Detailed structural analysis of \textit{a}-ZnON-21 and \textit{a}-ZnON-11 structures is provided in the supplementary information (SI-II).

\section{Electronic structure of \lowercase{\textit{a}}-Z\lowercase{n}ON and \lowercase{\textit{a}}-IGZO}
%\begin{figure*}[]
%\includegraphics[width=1.0 \linewidth]{./figures/final.pdf}
%%\captionsetup{justification=centering}
%\caption[]
%{(Color online) Electronic band structure and PDOS plots for (a) \textit{a}-ZnON-21 and (b) \textit{a}-ZnON-11 compositions and for (c) \textit{a}-IGZO-1114 and (d) \textit{a}-IGZO-2217 compositions. These plots are constructed for a representative structure from each type and composition. Colors used for different orbitals are shown below using a qualitative color bar.}	
%\label{fig.2}
%\end{figure*}

By nature, amorphous structures lack long range periodicity, making them difficult to be treated using DFT, which works best for periodic structures. However, DFT can still be employed, by making large unit-cells of amorphous materials with random distribution of atomic species. Electronic structure calculations were performed using plane-wave code Quantum Espresso with Perdew-Burke-Ernzerhof (PBE) ultrasoft pseudo-potentials \cite{38, 39, 40}, using a kinetic energy cut-off of 50 Ry, along with a 4x4x4 Monkhorst-Pack k-point grid for the relevant reciprocal space integration.

The representative pseudo-band structures (a-d) and the corresponding orbital-resolved partial electronic density of states (PDOS) (e-h), for the two compositions of \textit{a}-ZnON and \textit{a}-IGZO, are shown in Fig.\ref{fig.2}. All amorphous structures were generated in the same way as discussed earlier, except for \textit{a}-IGZO-2217* (2217 represents the atomic ratios of In, Ga , Zn and O) structure, which was obtained from the previous work of Divya \textit{et al.} \cite{48}. \textit{a}-IGZO-1114 (1114 represents the atomic ratio of In, Ga, Zn and O) structures were taken from our previous work\cite{37} and all the structures have densities in the range of 85\%-91\% of the theoretical densities of their crystalline counterparts. An apparent band gap at the gamma point shows the semiconducting nature of these amorphous structures. 
An important difference between the PDOS of \textit{a}-IGZO and \textit{a}-ZnON is that the valence band edge of former is made of oxygen-2\textit{p} orbitals (red), while the valence band edge of latter is made of nitrogen-2\textit{p} orbitals (blue), which is also consistent with observations reported in literature\cite{49, 50, 51, 52, 27, 29, 31}.
The conduction band edge of \textit{a}-ZnON is primarily made of Zn-4\textit{s} orbitals, while in \textit{a}-IGZO, the conduction band edge is dominated by In-5\textit{s} orbitals but significant contributions from other metal s-orbitals can also be seen from the insets in PDOS plots of \textit{a}-IGZO.

While most of the recent literature on \textit{a}-IGZO attribute overlapping In-5\textit{s} orbitals as the electron conduction path through the structure\cite{49, 50, 51, 52}, one of the earliest articles reported overlapping Zn-4\textit{s} orbitals as the conduction path in \textit{a}-IGZO\cite{53}.
\begin{figure*}[]
\includegraphics[width=0.85 \linewidth]{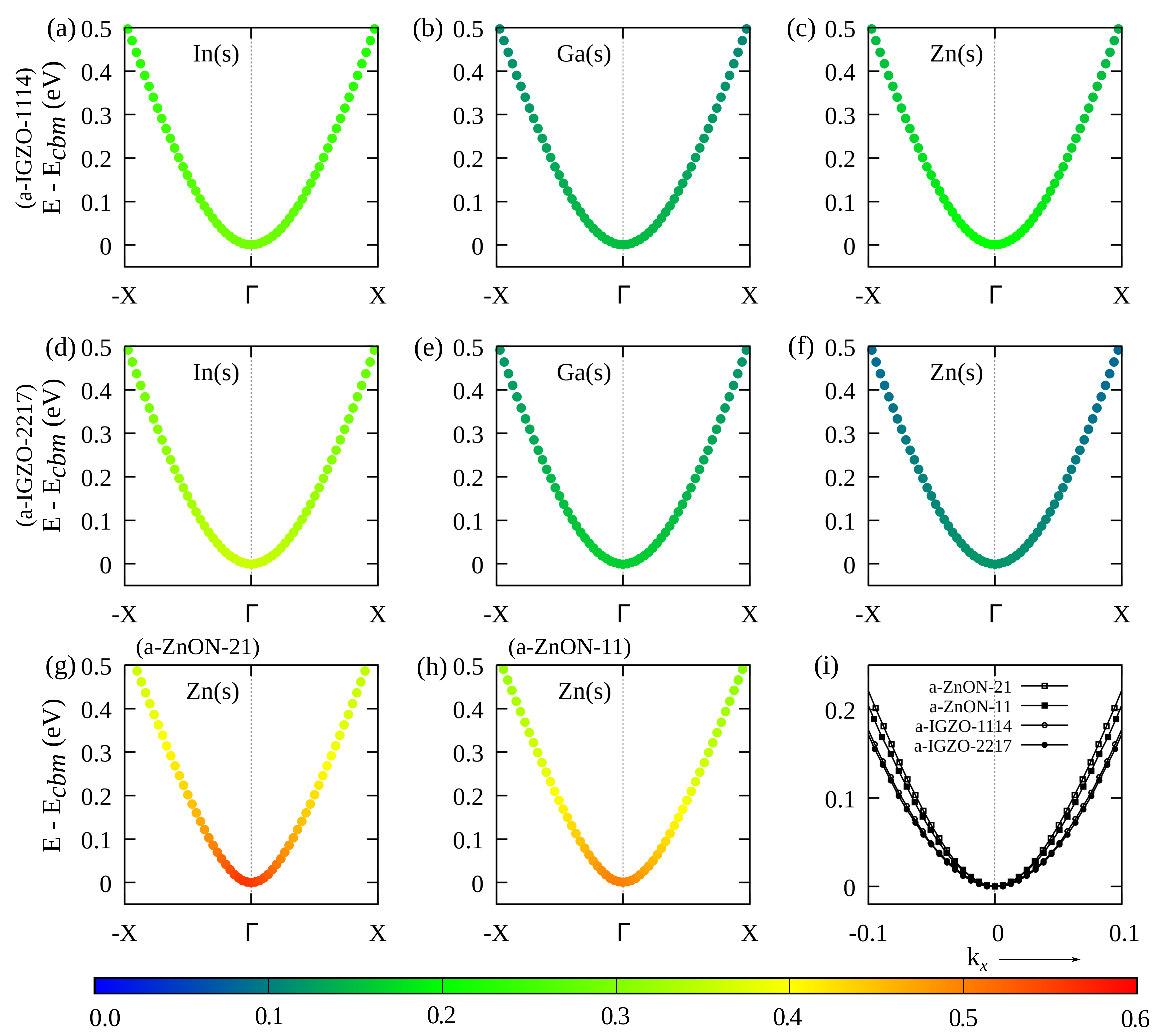}
%\captionsetup{justification=centering}
\caption[]
{(Color online) Orbital-resolved electronic band structures; for \textit{s}-orbitals of three metal cation (In, Ga and Zn respectively) (a), (b), (c) for \textit{a}-IGZO-1114 composition; (d), (e), (f) for \textit{a}-IGZO-2217 composition. The contribution towards CBM from \textit{s}-orbitals of Zn metal cation for (g) for \textit{a}-ZnON-21 and (h) \textit{a}-ZnON-11 compositions. (i) shows the difference in curvature of the CBM for \textit{a}-IGZO and \textit{a}-ZnON structures (Color bar shows the color gradient for fractional contribution from different orbitals under consideration).}	
\label{fig.3}
\end{figure*}
\begin{table*}[]
\centering
\begin{tabular}{ c | c | c | c | c | c | c | c }
%{ P{1.5cm}  P{1.5cm} | P{1.5cm}  P{1.5cm} | P{1.5cm}  P{1.5cm} | P{1.5cm}  P{1.5cm} }
%\hline
\hline
\multicolumn{2}{c|}{\textit{a}-ZnON-21} & \multicolumn{2}{c|}{\textit{a}-ZnON-11} & \multicolumn{2}{c|}{\textit{a}-IGZO-1114} & \multicolumn{2}{c}{\textit{a}-IGZO-2217} \\ 
\hline
Orbital &  PDOS \% & Orbital & PDOS \% & Orbital & PDOS \% & Orbital & PDOS \% \\
\hline
Zn-s  & 55.53 & Zn-s & 49.90 & In-s & 29.10 & In-s & 35.76 \\
%\hline
N-s & 14.79 & N-s & 21.09 & O-s & 22.78 & O-s & 22.23 \\ 
%\hline
O-s & 13.01 & O-s & 7.24 & Zn-s & 20.11 & Ga-s & 16.29 \\
%\hline
O-p & 3.59 & N-p & 4.62 & Ga-s & 14.92 & Zn-s & 11.71 \\
%\hline
N-p & 2.55 & O-p & 2.31 & O-p & 9.50 & O-p & 11.02 \\
%\hline
\hline
\end{tabular}
\caption[]
{Contributions of various orbitals towards the conduction band minimum (at the $\Gamma$-point) in two different compositions of \textit{a}-ZnON and \textit{a}-IGZO. These values are calculated for a representative structure from each type and composition.}
\label{table.1}
\end{table*}
In our calculations for \textit{a}-IGZO, we see significant contributions form Zn-4\textit{s} and Ga-4\textit{s} orbitals to the CBM in PDOS plots (Fig.\ref{fig.2} g-h). To quantify the contributions of various metal \textit{s}-orbitals, we calculated the percentage contribution based on the projection of wavefunctions on atomic basis formed as a linear combination of atomic orbitals, using the method implemented in QE code\cite{58}.
The spilling parameter, which is indicative of mismatch between plane-wave wavefunctions and its projection onto the atomic basis, was found to be very small (0.0019 - 0.0026) for all compositions. In Fig.\ref{fig.3}, we plot orbital resolved (metal-\textit{s} orbitals only) band structures in the vicinity of CBM and in Table.\ref{table.1}, percentage contributions of metal-\textit{s} orbitals towards total density of states at the CBM are listed. %Note that the values reported in Table.\ref{table.1} are a qualitative estimate of the contributions of different metal cations towards conduction and its dependence on composition.
From Table.\ref{table.1}, we see that the contribution of Zn-\textit{s} orbitals is about 50\% at the CBM for \textit{a}-ZnON structures. For \textit{a}-IGZO, contributions from different metal-\textit{s} orbitals are comparable and the contributions from Zn-\textit{s} and Ga-\textit{s} orbitals (in comparison to In-\textit{s} orbitals) cannot be ignored. The contributions towards CBM from different metal-\textit{s} orbitals also follow chemical composition, as evident from the higher Zn-\textit{s} contribution in \textit{a}-IGZO-1114 in comparison to \textit{a}-IGZO-2217. From the orbital-resolved band structure plot in Fig.\ref{fig.3} also, we can see that the contributions of different metal-\textit{s} orbitals, in the vicinity of CBM, are comparable for \textit{a}-IGZO compositions. Hence, we argue that the electronic transport in \textit{a}-IGZO should be through overlapping \textit{s}-orbitals of different metal cations rather than only In-5\textit{s} orbitals, as cited in several papers \cite{49, 50, 51, 52}. For \textit{a}-ZnON compositions, the highest contribution towards CBM from Zn-4\textit{s} orbitals is also clearly evident here. 
A comparison of the curvature at CBM for \textit{a}-IGZO and \textit{a}-ZnON compositions (thus comparing the effective mass of the two qualitatively) is shown in Fig.\ref{fig.3}(i). 
From this composite conduction band plot, the higher curvature at CBM for \textit{a}-ZnON is apparent, signifying its lower electron effective mass, which can be related to the higher electron mobilities observed for \textit{a}-ZnON semiconductors.

In \textit{a}-IGZO, electron mobility also depends on the carrier concentration (due to the percolation mechanism of electron transport\cite{60, 61}) and increasing In content in the structure increases electron concentration (due to increase in oxygen-vacancies), which explains the observed higher mobility. While increasing Ga content, electron carrier concentration decreases and hence the measured mobilities are comparatively low \cite{14, 60, 61}. It is very difficult to deconvolute the role of cations in electron transport from the measured electron mobilities, without including the effect of carrier concentration due to oxygen vacancies. Though, the formation of oxygen vacancies in ionic oxide semiconductors is mainly governed by the thermodynamics of material system, it also heavily depends on the processing conditions, such as partial pressure of oxygen during thin film deposition (sputtering) and/or subsequent annealing.
In the next section, we discuss electron transport, at atomic scale, between empty outer \textit{s}-orbital of two adjacent metal cations in terms of orbital overlap integral to understand the role of metal cations in forming the electron conduction path through these amorphous semiconductors.

\section{Orbital overlap integral and electronic conduction}
In the previous discussion, we argued that the electronic conduction path must be through overlapping \textit{s}-orbitals of different cations in \textit{a}-IGZO and overlapping Zn-4\textit{s} orbitals in \textit{a}-ZnON. Therefore, the ease of electron transport (electron mobility) must depend on the nature and extent of overlap between these orbitals. 
In the following discussion, we attempt to quantify the overlap between outermost \textit{s}-orbitals of different metal cations in these amorphous semiconductors and relate the overlap to physically verifiable quantities such as electron mobilities. Orbital overlap integral (OI) between different orbitals is a measure of overlap between single electron wavefunctions in these orbitals. Orita \textit{et al.} \cite{53} argued that for electron transport in oxide semiconductors, the orbital overlap integral (OI) for metal-\textit{s} orbitals should be greater than 0.4 and the fraction of metal cations forming the conduction path should be above a percolation threshold of 20\%. From the second condition, it is perceptible that in \textit{a}-IGZO compositions (1114 and 2217), the electron conduction path can not be formed by only a single type of metal cation and hence, in \textit{a}-IGZO, the conduction path must be through overlapping \textit{s}-orbitals of different metal cations, while in \textit{a}-ZnON, the conduction path is through overlapping Zn-4\textit{s} orbitals. To understand the contribution of different metal-\textit{s} orbitals and their overlap, we calculated overlap integrals between outermost \textit{s} orbitals of different pairs of metal ions, for both compositions of \textit{a}-IGZO and \textit{a}-ZnON. We followed Mulliken’s formulation for calculating overlap integrals for Slater-type atomic orbitals (STOs) \cite{54}. The formulation used in our calculations is given in supplementary information (SI-III). The numerical integrations were carried out using MATHEMATICA \cite{55}.
Different structures for the two compositions of \textit{a}-IGZO and two compositions of \textit{a}-ZnON were analyzed to calculate pairwise orbital overlap integrals. A cut-off distance of 4 {\AA} was chosen to be the radius of the interaction sphere around each metal cation and all other cations in the interaction sphere were assumed to have overlapping \textit{s} orbital with the metal cation in the center.  Table.\ref{table.2} shows the number of metal ion pairs, their average distances R (\AA) and the orbital overlap integral (OI) for outermost \textit{s} orbitals, calculated at average distances, for the two compositions of \textit{a}-IGZO and \textit{a}-ZnON. The total contribution from a particular metal ion pair per unit volume (Number of Pairs * OI / volume) is also shown in Table.\ref{table.2}, which can be compared between different compositions and the sum over all pairs is considered for \textit{a}-IGZO. We notice a marked difference in the calculated overlap integrals for pairs of similar cations (Zn-Zn, In-In and Ga-Ga) with similar average cation-cation distances, compared to the values reported by Orita \textit{et al.} \cite{53} and the orbital overlap between any pair is also less than 0.4, which was proposed as a criteria for forming electron transport path. 
We ascribe this deviation to the slight difference in way the orbital overlap integrals are calculated. The orbital overlap integral values for different metal-cation pairs range from 0.13 to 0.30, as shown in Table.\ref{table.2}. All these cation-pairs should form the conducting pathways for electron transport, however, with varying ease of electron transport between two overlapping orbitals, with higher overlap integral providing less resistance to electron transport.

\begin{table*}[]
%\captionsetup{justification=centering}
\centering
\begin{tabular}{ c | c | c | c | c | c | c | c | c}
\multicolumn{2}{c}{\textbf{(a)}} \\
%\hline
\hline
 & \multicolumn{4}{c|}{\textit{a}-IGZO-1114} & \multicolumn{4}{c}{\textit{a}-IGZO-2217} \\ 
  & \multicolumn{4}{c|}{Voume=1148.19645 {\AA}$^3$} & \multicolumn{4}{c}{Voume=1152.70734 {\AA}$^3$} \\ 
\hline
Pair & No. of Pairs & R ({\AA}) & Overlap Integral & OI*Pairs/Volume & No. of Pairs & R ({\AA}) & Overlap Integral & OI*Pairs/Volume \\
\hline
Zn-Zn  & 16 & 3.44 & 0.256 & 0.00357 & 3 & 3.48 & 0.248 & 0.00064  \\
%\hline
In-In & 15 & 3.60 & 0.186 & 0.00243 & 25 & 3.52 & 0.200 & 0.00434 \\ 
%\hline
Ga-Ga & 14 & 3.43 & 0.134 & 0.00163 & 22 & 3.31 & 0.154 & 0.00294 \\
%\hline
In-Zn & 39 & 3.48 & 0.226 & 0.00768 & 32 & 3.51 & 0.222 & 0.00615 \\
%\hline
In-Ga & 45 & 3.43 & 0.174 & 0.00681 & 56 & 3.53 & 0.157 & 0.00761 \\
%\hline
Zn-Ga & 39 & 3.28 & 0.222 & 0.00756 & 28 & 3.36 & 0.206 & 0.00500 \\
\hline 
\textbf{Sum} & & & & 0.02966 & & & & 0.02669 \\
\hline 
\end{tabular}

\bigskip 

\begin{tabular}{ c | c | c | c | c | c | c | c | c }
\multicolumn{2}{c}{\textbf{(b)}} \\
\hline 
 & \multicolumn{4}{c|}{\textit{a}-ZnON-21} & \multicolumn{4}{c}{\textit{a}-ZnON-11} \\ 
  & \multicolumn{4}{c|}{Volume=1081.7498 {\AA}$^3$} & \multicolumn{4}{c}{Volume=1506.9619 {\AA}$^3$} \\
\hline
Pair & No. of Pairs & R ({\AA}) & Overlap Integral & Pairs*OI/Volume & No. of Pairs & R ({\AA}) & Overlap Integral & Pairs*OI/Volume \\
\hline
Zn-Zn  & 232 & 3.27 & 0.291 & 0.06231 & 337 & 3.23 & 0.299 & 0.06693  \\
\hline 
%\hline
\end{tabular}

\caption[]
{Pair-wise orbital overlap integral for metal cation pairs in case of (a) \textit{a}-IGZO, two compositions and, (b) \textit{a}-ZnON, two compositions, at their average distances within a cut off radius of  4 {\AA}, selected to be the radius of the interaction sphere.}
\label{table.2}
\end{table*}

The sum of all pairwise orbital overlap integrals, normalized to volume (total normalized orbital overlap integral), gives us a parameter to compare electron transport among different structures and compositions. It should be noted that the orbital overlap integral would strongly depend on the density of the amorphous structure, as with increasing density the average distances between metal cation pairs would decrease. For two different compositions of \textit{a}-IGZO, the total normalized orbital overlap integrals are very similar with slightly higher value for \textit{a}-IGZO-1114 composition compared to \textit{a}-IGZO-2217. From Table.\ref{table.2}(a), it can be seen that the number of pairs for mix cations is much higher in comparison to the number of pairs of similar cations for \textit{a}-IGZO. The value of overlap integral is found to be the highest for Zn-Zn pairs, but since the number of pairs are limited, this overlap cannot be solely responsible for the electronic conduction in the material. From the table, it is also evident that for \textit{a}-IGZO-1114 composition, the main contribution to overall electron transport comes from electron transport between In-Zn, Zn-Ga and In-Ga cation pairs, and for \textit{a}-IGZO-2217 composition, the main contribution is from electron transport between In-Ga and In-Zn metal-cation pairs. These results are also consistent with our observation of the contribution of different metal \textit{s}-orbitals in conduction band of \textit{a}-IGZO (Fig.\ref{fig.3} and Table.\ref{table.1}). For single cation amorphous structures, such as \textit{a}-ZnON, electron transport path is always through overlapping \textit{s}-orbitals of Zn-Zn metal-cation pairs. Since there is only one type of metal-cations present in this system and the average distance between Zn-Zn ions is smaller than pairs of any two cations in \textit{a}-IGZO, the orbital overlap is higher for \textit{a}-ZnON in comparison to \textit{a}-IGZO. Also, owing to the large number of such pairs with higher orbital overlap, the total normalized orbital overlap integral for \textit{a}-ZnON is much higher (more than a factor of 2) compared to \textit{a}-IGZO and hence, we may expect the electron transport to be much easier in \textit{a}-ZnON. From literature \cite{28, 33, 13, 14, 15, 16, 17}, we know that the measured Hall mobilities for \textit{a}-ZnON are much higher compared to \textit{a}-IGZO, which is consistent with higher values of normalized orbital overlap integrals, observed from our calculations, for \textit{a}-ZnON compared to \textit{a}-IGZO. 

Although we cannot directly relate the total overlap integral with measured values of hall mobilities, there is an apparent correlation between the two. Carrier mobilities in any material depend on the effective mass and the scattering time constants. The scattering mechanisms in amorphous semiconductors are not clearly understood and often difficult to measure or calculate. The carrier effective mass, on the other hand, can be estimated from the electronic structure of material and can give us valuable insight regarding carrier transport. From the pseudo band structures of \textit{a}-IGZO and \textit{a}-ZnON in Fig.\ref{fig.2}, the bottom of the conduction band appears to be symmetric around $\Gamma$ point and effective mass is calculated by fitting a parabola in small \textit{k}-range of $\pm$ 0.1 {\AA}$^{-1}$ in $\Gamma$-X direction at the minima and taking the inverse of the curvature ($\frac{1}{m^*}=\frac{1}{\hbar^2}(\frac{\partial^2E}{\partial k^2})$). Since the structures are amorphous, the effective masses in three different directions might be slightly different, owing to different arrangements of ions in the three directions in real space. 
\begin{figure}[]
\includegraphics[width=1.0 \linewidth]{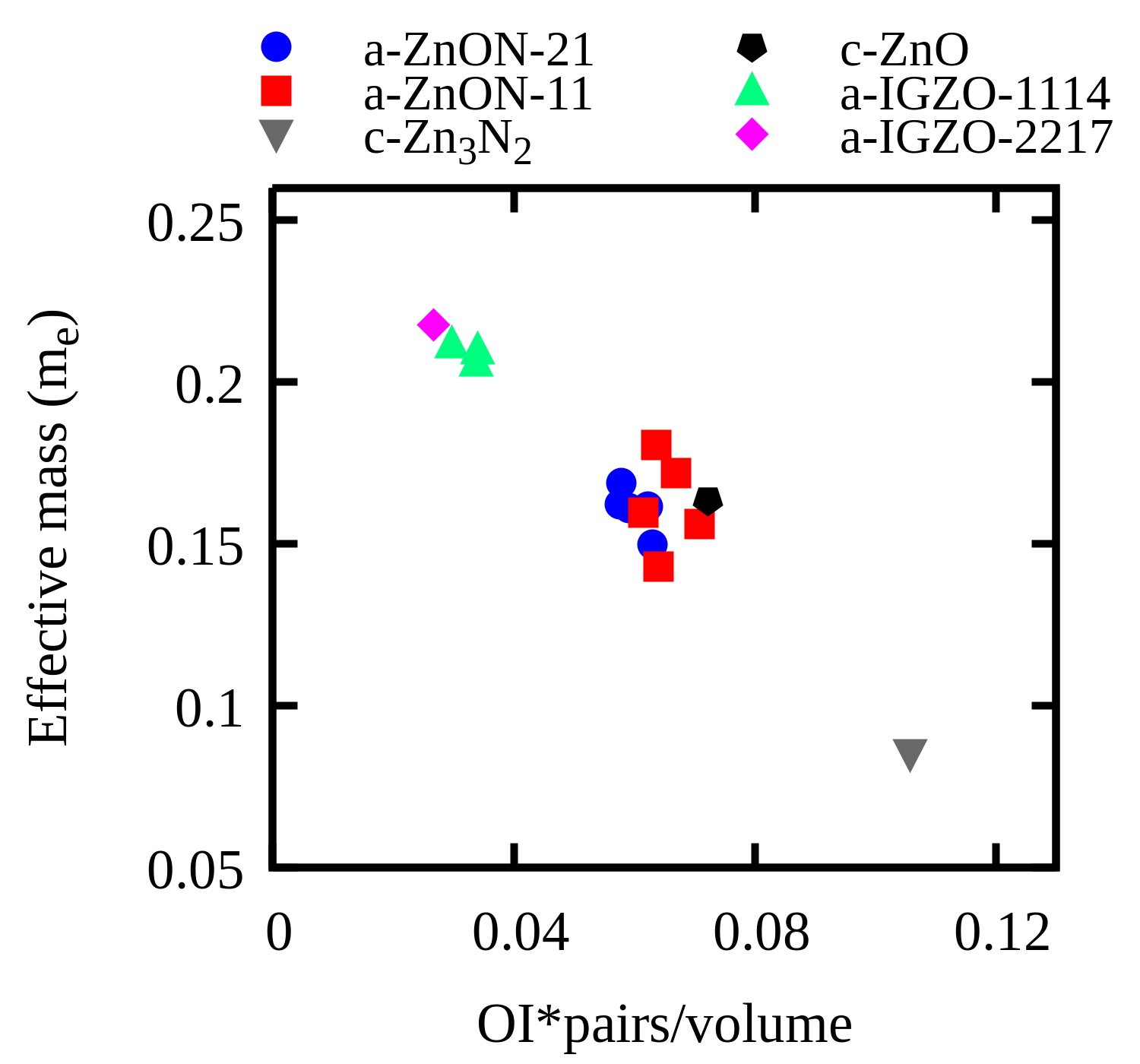}
%\captionsetup{justification=centering}
\caption[]
{(Color online) Correlation between the effective mass and the total normalized orbital overlap integral (OI * Number of Pairs/volume) for different structures of \textit{a}-IGZO-1114, \textit{a}-IGZO-2217*, \textit{a}-ZnON-21 and \textit{a}-ZnON-11 compositions. Same is also shown for \textit{c}-ZnO and \textit{c}-Zn$_3$N$_2$.}	
\label{fig.4}
\end{figure}
We took the average by calculating the effective masses in all three ($\Gamma$-X, $\Gamma$-Y and $\Gamma$-Z) directions. As shown in Table.\ref{table.2}, for \textit{a}-IGZO-1114 and \textit{a}-IGZO-2217 structures, the electron effective mass was estimated to be 0.21 m$_e$ and 0.22 m$_e$ respectively. While these two values are nearly same, they are small compared to the measured values of 0.34 m$_e$ \cite{16} for electron effective mass in \textit{a}-IGZO but in agreement with calculated values reported in the Ref.\cite{15}. For \textit{a}-ZnON-11 and \textit{a}-ZnON-21 structures, as shown in Table.\ref{table.2}, the estimated electron effective masses are 0.17 m$_e$ and 0.16 m$_e$ respectively, which also match closely with previously reported theoretical value of 0.19 m$_e$ for electron effective mass in \textit{a}-ZnON \cite{29}. The smaller electron effective mass observed in \textit{a}-ZnON compared to \textit{a}-IGZO, suggest faster electron transport in \textit{a}-ZnON. Our earlier discussion on total normalized orbital overlap integral also had a similar observation and hence we expect to see some correlation between two calculated values; the effective mass and the total normalized orbital overlap integral\cite{59}. The difference in calculated electron effective mass between a-IGZO and a-ZnON is not enough to explain the differences in observed electron mobilities in these two materials. The difference in total normalized OI explains the observed difference in mobilities to a better extent but still may not be adequate\cite{59}. 
To explore this correlation, we plot (Fig.\ref{fig.4}) the effective mass and the total normalized orbital overlap integral for different structures of \textit{a}-IGZO-1114 (three structures), \textit{a}-IGZO-2217* (one structure), \textit{a}-ZnON-11 (five structures) and \textit{a}-ZnON-21 (five structures) compositions. For comparison, we also plot the estimated effective mass and calculated orbital overlap integrals for \textit{c}-ZnO and \textit{c}-Zn$_3$N$_2$. From the plot, we can see a strong correlation between effective mass, calculated using \textit{ab initio} method and the total normalized orbital overlap integral which is calculated from the structural information. Hence, we can use the normalized orbital overlap integral as an estimate for the carrier effective mass and the carrier mobility, for comparison between amorphous structures.
The measured Hall mobilities strongly depend on deposition parameters and electron concentration and hence, a direct correlation between measured properties (Hall mobility) with calculated values (effective mass or total normalized OI) can only be qualitative, representing a trend to facilitate further exploration rather than being definitive. Among the two calculated properties; the effective mass and total normalized OI; the total normalized OI seems to relate better with measured mobilities.

\section{Conclusions}
{
A multi-anion approach towards amorphous semiconductors has distinct advantages over multi-cation based amorphous oxide semiconductors. Comparison of electronic structures of the two materials systems (\textit{a}-ZnON and \textit{a}-IGZO) highlights the role of cations and anions in the electronic transport. While the electronic conduction in \textit{a}-IGZO is found to be through overlapping empty \textit{s}-orbitals of mixed cations and so it also depends on composition (ratio of cations), in \textit{a}-ZnON, the electronic conduction path is through overlapping Zn-\textit{s} orbitals and is independent of composition (ratio of anions). The ease of conduction in these amorphous structure can be estimated in terms of extent of overlap between empty \textit{s} orbitals of metal cations.
The total normalized orbital overlap integral is shown to have a direct correlation with calculated carrier effective mass, this correlation can be used as a qualitative estimate for the observed electron mobilities. Effective mass calculations based on \textit{ab initio} DFT calculation of electronic structure are computationally expensive, on the other hand, the calculation of the total normalized  orbital overlap integral as an estimate for carrier effective mass can be done on for many different amorphous structures or structural averages from many amorphous structures can be calculated with very small computational expense. Calculation of the pairwise orbital overlap integrals for different metal cation pairs also provides an insight into the role of individual metal cations in forming the conduction path in multi-cation amorphous structures which is not straightforward from the DFT calculations. While single structure based DFT calculation are limited in their scope, calculations based on many structures or averages of many structures, such as generated by evolutionary algorithm based codes, should give us better insight into the average properties and role of individual cations/anions in these materials and can be used effectively while predicting properties of new compositions of amorphous semiconductors.
}

\begin{acknowledgments}
{
Authors would like to acknowledge support from HPC, IIT Kanpur for use of their computing facility in the present work. J. Srivastava acknowledges financial support from MHRD, Govt. of India and A. Gaur acknowledges support from MeitY, Govt. of India under Visvesvaraya YFRF program. Authors also thank Divya et. al. for providing a-IGZO-2217 structure for comparative studies.
}
\end{acknowledgments}

%\appendix

%\section{Appendixes}

\nocite{*}
\bibliography{references}% Produces the bibliography via BibTeX.

\begin{thebibliography}{58}%
\makeatletter
\providecommand \@ifxundefined [1]{%
 \@ifx{#1\undefined}
}%
\providecommand \@ifnum [1]{%
 \ifnum #1\expandafter \@firstoftwo
 \else \expandafter \@secondoftwo
 \fi
}%
\providecommand \@ifx [1]{%
 \ifx #1\expandafter \@firstoftwo
 \else \expandafter \@secondoftwo
 \fi
}%
\providecommand \natexlab [1]{#1}%
\providecommand \enquote  [1]{``#1''}%
\providecommand \bibnamefont  [1]{#1}%
\providecommand \bibfnamefont [1]{#1}%
\providecommand \citenamefont [1]{#1}%
\providecommand \href@noop [0]{\@secondoftwo}%
\providecommand \href [0]{\begingroup \@sanitize@url \@href}%
\providecommand \@href[1]{\@@startlink{#1}\@@href}%
\providecommand \@@href[1]{\endgroup#1\@@endlink}%
\providecommand \@sanitize@url [0]{\catcode `\\12\catcode `\$12\catcode
  `\&12\catcode `\#12\catcode `\^12\catcode `\_12\catcode `\%12\relax}%
\providecommand \@@startlink[1]{}%
\providecommand \@@endlink[0]{}%
\providecommand \url  [0]{\begingroup\@sanitize@url \@url }%
\providecommand \@url [1]{\endgroup\@href {#1}{\urlprefix }}%
\providecommand \urlprefix  [0]{URL }%
\providecommand \Eprint [0]{\href }%
\providecommand \doibase [0]{http://dx.doi.org/}%
\providecommand \selectlanguage [0]{\@gobble}%
\providecommand \bibinfo  [0]{\@secondoftwo}%
\providecommand \bibfield  [0]{\@secondoftwo}%
\providecommand \translation [1]{[#1]}%
\providecommand \BibitemOpen [0]{}%
\providecommand \bibitemStop [0]{}%
\providecommand \bibitemNoStop [0]{.\EOS\space}%
\providecommand \EOS [0]{\spacefactor3000\relax}%
\providecommand \BibitemShut  [1]{\csname bibitem#1\endcsname}%
\let\auto@bib@innerbib\@empty
%</preamble>
\bibitem [{\citenamefont {Momma}\ and\ \citenamefont {Izumi}(2011)}]{46}%
  \BibitemOpen
  \bibfield  {author} {\bibinfo {author} {\bibfnamefont {K.}~\bibnamefont
  {Momma}}\ and\ \bibinfo {author} {\bibfnamefont {F.}~\bibnamefont {Izumi}},\
  }\bibfield  {title} {\enquote {\bibinfo {title} {{{\it VESTA3} for
  three-dimensional visualization of crystal, volumetric and morphology
  data}},}\ }\href {\doibase 10.1107/S0021889811038970} {\bibfield  {journal}
  {\bibinfo  {journal} {Journal of Applied Crystallography}\ }\textbf {\bibinfo
  {volume} {44}},\ \bibinfo {pages} {1272--1276} (\bibinfo {year}
  {2011})}\BibitemShut {NoStop}%
\bibitem [{\citenamefont {Powell}(1983)}]{1}%
  \BibitemOpen
  \bibfield  {author} {\bibinfo {author} {\bibfnamefont {M.~J.}\ \bibnamefont
  {Powell}},\ }\bibfield  {title} {\enquote {\bibinfo {title} {{Charge trapping
  instabilities in amorphous silicon-silicon nitride thin-film transistors}},}\
  }\href {\doibase 10.1063/1.94399} {\bibfield  {journal} {\bibinfo  {journal}
  {Applied Physics Letters}\ }\textbf {\bibinfo {volume} {43}},\ \bibinfo
  {pages} {597--599} (\bibinfo {year} {1983})}\BibitemShut {NoStop}%
\bibitem [{\citenamefont {Libsch}\ and\ \citenamefont {Kanicki}(1993)}]{2}%
  \BibitemOpen
  \bibfield  {author} {\bibinfo {author} {\bibfnamefont {F.~R.}\ \bibnamefont
  {Libsch}}\ and\ \bibinfo {author} {\bibfnamefont {J.}~\bibnamefont
  {Kanicki}},\ }\bibfield  {title} {\enquote {\bibinfo {title} {{Bias stress
  induced stretched exponential time dependence of charge injection and
  trapping in amorphous thin film transistors}},}\ }\href {\doibase
  10.1063/1.108709} {\bibfield  {journal} {\bibinfo  {journal} {Applied Physics
  Letters}\ }\textbf {\bibinfo {volume} {62}},\ \bibinfo {pages} {1286--1288}
  (\bibinfo {year} {1993})}\BibitemShut {NoStop}%
\bibitem [{\citenamefont {Street}(2009)}]{3}%
  \BibitemOpen
  \bibfield  {author} {\bibinfo {author} {\bibfnamefont {R.~A.}\ \bibnamefont
  {Street}},\ }\bibfield  {title} {\enquote {\bibinfo {title} {{Thin-Film
  Transistors}},}\ }\href {\doibase 10.1002/adma.200803211} {\bibfield
  {journal} {\bibinfo  {journal} {Advanced Materials}\ }\textbf {\bibinfo
  {volume} {21}},\ \bibinfo {pages} {2007--2022} (\bibinfo {year}
  {2009})}\BibitemShut {NoStop}%
\bibitem [{\citenamefont {Street}(2005)}]{4}%
  \BibitemOpen
  \bibfield  {author} {\bibinfo {author} {\bibfnamefont {R.}~\bibnamefont
  {Street}},\ }\href {https://books.google.co.in/books?id=zQMwsMt4VmoC} {\emph
  {\bibinfo {title} {{Hydrogenated Amorphous Silicon}}}},\ Cambridge Solid
  State Science Series\ (\bibinfo  {publisher} {Cambridge University Press},\
  \bibinfo {year} {2005})\BibitemShut {NoStop}%
\bibitem [{\citenamefont {Jang}, \citenamefont {Choi},\ and\ \citenamefont
  {Cheon}()}]{56}%
  \BibitemOpen
  \bibfield  {author} {\bibinfo {author} {\bibfnamefont {J.}~\bibnamefont
  {Jang}}, \bibinfo {author} {\bibfnamefont {M.~H.}\ \bibnamefont {Choi}}, \
  and\ \bibinfo {author} {\bibfnamefont {J.~H.}\ \bibnamefont {Cheon}},\
  }\bibfield  {title} {\enquote {\bibinfo {title} {{77.1: Invited Paper: TFT
  Technologies for Flexible Displays}},}\ }\href {\doibase 10.1889/1.3499860}
  {\bibfield  {journal} {\bibinfo  {journal} {SID Symposium Digest of Technical
  Papers}\ }\textbf {\bibinfo {volume} {41}},\ \bibinfo {pages}
  {1143--1146}}\BibitemShut {NoStop}%
\bibitem [{\citenamefont {Kim}\ \emph {et~al.}()\citenamefont {Kim},
  \citenamefont {Jeong}, \citenamefont {Chung},\ and\ \citenamefont {Mo}}]{57}%
  \BibitemOpen
  \bibfield  {author} {\bibinfo {author} {\bibfnamefont {H.~D.}\ \bibnamefont
  {Kim}}, \bibinfo {author} {\bibfnamefont {J.~K.}\ \bibnamefont {Jeong}},
  \bibinfo {author} {\bibfnamefont {H.-J.}\ \bibnamefont {Chung}}, \ and\
  \bibinfo {author} {\bibfnamefont {Y.-G.}\ \bibnamefont {Mo}},\ }\bibfield
  {title} {\enquote {\bibinfo {title} {{22.1: Invited Paper: Technological
  Challenges for Large-Size AMOLED Display}},}\ }\href {\doibase
  10.1889/1.3069649} {\bibfield  {journal} {\bibinfo  {journal} {SID Symposium
  Digest of Technical Papers}\ }\textbf {\bibinfo {volume} {39}},\ \bibinfo
  {pages} {291--294}}\BibitemShut {NoStop}%
\bibitem [{\citenamefont {Hahn}(1951)}]{5}%
  \BibitemOpen
  \bibfield  {author} {\bibinfo {author} {\bibfnamefont {E.~E.}\ \bibnamefont
  {Hahn}},\ }\bibfield  {title} {\enquote {\bibinfo {title} {{Some Electrical
  Properties of Zinc Oxide Semiconductor}},}\ }\href {\doibase
  10.1063/1.1700063} {\bibfield  {journal} {\bibinfo  {journal} {Journal of
  Applied Physics}\ }\textbf {\bibinfo {volume} {22}},\ \bibinfo {pages}
  {855--863} (\bibinfo {year} {1951})}\BibitemShut {NoStop}%
\bibitem [{\citenamefont {Janotti}\ and\ \citenamefont {de~Walle}(2009)}]{6}%
  \BibitemOpen
  \bibfield  {author} {\bibinfo {author} {\bibfnamefont {A.}~\bibnamefont
  {Janotti}}\ and\ \bibinfo {author} {\bibfnamefont {C.~G.~V.}\ \bibnamefont
  {de~Walle}},\ }\bibfield  {title} {\enquote {\bibinfo {title} {{Fundamentals
  of zinc oxide as a semiconductor}},}\ }\href
  {http://stacks.iop.org/0034-4885/72/i=12/a=126501} {\bibfield  {journal}
  {\bibinfo  {journal} {Reports on Progress in Physics}\ }\textbf {\bibinfo
  {volume} {72}},\ \bibinfo {pages} {126501} (\bibinfo {year}
  {2009})}\BibitemShut {NoStop}%
\bibitem [{\citenamefont {Navamathavan}, \citenamefont {Choi},\ and\
  \citenamefont {Park}(2009)}]{7}%
  \BibitemOpen
  \bibfield  {author} {\bibinfo {author} {\bibfnamefont {R.}~\bibnamefont
  {Navamathavan}}, \bibinfo {author} {\bibfnamefont {C.~K.}\ \bibnamefont
  {Choi}}, \ and\ \bibinfo {author} {\bibfnamefont {S.-J.}\ \bibnamefont
  {Park}},\ }\bibfield  {title} {\enquote {\bibinfo {title} {{Electrical
  properties of ZnO-based bottom-gate thin film transistors fabricated by using
  radio frequency magnetron sputtering}},}\ }\href {\doibase
  https://doi.org/10.1016/j.jallcom.2008.08.039} {\bibfield  {journal}
  {\bibinfo  {journal} {Journal of Alloys and Compounds}\ }\textbf {\bibinfo
  {volume} {475}},\ \bibinfo {pages} {889 -- 892} (\bibinfo {year}
  {2009})}\BibitemShut {NoStop}%
\bibitem [{\citenamefont {Lee}\ \emph {et~al.}(2012)\citenamefont {Lee},
  \citenamefont {Shih}, \citenamefont {Chen}, \citenamefont {Huang},
  \citenamefont {Gan}, \citenamefont {Shih}, \citenamefont {Qiu},\ and\
  \citenamefont {Shih}}]{8}%
  \BibitemOpen
  \bibfield  {author} {\bibinfo {author} {\bibfnamefont {M.-H.}\ \bibnamefont
  {Lee}}, \bibinfo {author} {\bibfnamefont {C.-C.}\ \bibnamefont {Shih}},
  \bibinfo {author} {\bibfnamefont {J.-S.}\ \bibnamefont {Chen}}, \bibinfo
  {author} {\bibfnamefont {W.-M.}\ \bibnamefont {Huang}}, \bibinfo {author}
  {\bibfnamefont {F.-Y.}\ \bibnamefont {Gan}}, \bibinfo {author} {\bibfnamefont
  {Y.-C.}\ \bibnamefont {Shih}}, \bibinfo {author} {\bibfnamefont {C.~X.}\
  \bibnamefont {Qiu}}, \ and\ \bibinfo {author} {\bibfnamefont {I.-S.}\
  \bibnamefont {Shih}},\ }\bibfield  {title} {\enquote {\bibinfo {title}
  {{15.4: Excellent Performance of Indium-Oxide-Based Thin-Film Transistors by
  DC Sputtering}},}\ }\href {\doibase 10.1889/1.3256729} {\bibfield  {journal}
  {\bibinfo  {journal} {SID Symposium Digest of Technical Papers}\ }\textbf
  {\bibinfo {volume} {40}},\ \bibinfo {pages} {191--193} (\bibinfo {year}
  {2012})}\BibitemShut {NoStop}%
\bibitem [{\citenamefont {Jin}\ \emph {et~al.}(2014)\citenamefont {Jin},
  \citenamefont {Luo}, \citenamefont {Bao}, \citenamefont {Brox-Nilsen},
  \citenamefont {Potter},\ and\ \citenamefont {Song}}]{9}%
  \BibitemOpen
  \bibfield  {author} {\bibinfo {author} {\bibfnamefont {J.}~\bibnamefont
  {Jin}}, \bibinfo {author} {\bibfnamefont {Y.}~\bibnamefont {Luo}}, \bibinfo
  {author} {\bibfnamefont {P.}~\bibnamefont {Bao}}, \bibinfo {author}
  {\bibfnamefont {C.}~\bibnamefont {Brox-Nilsen}}, \bibinfo {author}
  {\bibfnamefont {R.}~\bibnamefont {Potter}}, \ and\ \bibinfo {author}
  {\bibfnamefont {A.}~\bibnamefont {Song}},\ }\bibfield  {title} {\enquote
  {\bibinfo {title} {{Tuning the electrical properties of ZnO thin-film
  transistors by thermal annealing in different gases}},}\ }\href {\doibase
  https://doi.org/10.1016/j.tsf.2013.12.004} {\bibfield  {journal} {\bibinfo
  {journal} {Thin Solid Films}\ }\textbf {\bibinfo {volume} {552}},\ \bibinfo
  {pages} {192 -- 195} (\bibinfo {year} {2014})}\BibitemShut {NoStop}%
\bibitem [{\citenamefont {Kim}\ \emph {et~al.}(2014)\citenamefont {Kim},
  \citenamefont {Meng}, \citenamefont {Lee}, \citenamefont {Yu}, \citenamefont
  {Yoo}, \citenamefont {Kang},\ and\ \citenamefont {Jo}}]{10}%
  \BibitemOpen
  \bibfield  {author} {\bibinfo {author} {\bibfnamefont {J.}~\bibnamefont
  {Kim}}, \bibinfo {author} {\bibfnamefont {J.}~\bibnamefont {Meng}}, \bibinfo
  {author} {\bibfnamefont {D.}~\bibnamefont {Lee}}, \bibinfo {author}
  {\bibfnamefont {M.}~\bibnamefont {Yu}}, \bibinfo {author} {\bibfnamefont
  {D.}~\bibnamefont {Yoo}}, \bibinfo {author} {\bibfnamefont {D.~W.}\
  \bibnamefont {Kang}}, \ and\ \bibinfo {author} {\bibfnamefont
  {J.}~\bibnamefont {Jo}},\ }\bibfield  {title} {\enquote {\bibinfo {title}
  {{ZnO Thin-Film Transistor Grown by rf Sputtering Using Carbon Dioxide and
  Substrate Bias Modulation}},}\ }\href {\doibase
  https://doi.org/10.1155/2014/709018.} {\bibfield  {journal} {\bibinfo
  {journal} {Journal of Nanomaterials}\ }\textbf {\bibinfo {volume} {2014}},\
  \bibinfo {pages} {7} (\bibinfo {year} {2014})}\BibitemShut {NoStop}%
\bibitem [{\citenamefont {Hosono}, \citenamefont {Yasukawa},\ and\
  \citenamefont {Kawazoe}(1996)}]{11}%
  \BibitemOpen
  \bibfield  {author} {\bibinfo {author} {\bibfnamefont {H.}~\bibnamefont
  {Hosono}}, \bibinfo {author} {\bibfnamefont {M.}~\bibnamefont {Yasukawa}}, \
  and\ \bibinfo {author} {\bibfnamefont {H.}~\bibnamefont {Kawazoe}},\
  }\bibfield  {title} {\enquote {\bibinfo {title} {{Novel oxide amorphous
  semiconductors: transparent conducting amorphous oxides}},}\ }\href {\doibase
  https://doi.org/10.1016/0022-3093(96)00367-5} {\bibfield  {journal} {\bibinfo
   {journal} {Journal of Non-Crystalline Solids}\ }\textbf {\bibinfo {volume}
  {203}},\ \bibinfo {pages} {334 -- 344} (\bibinfo {year} {1996})},\ \bibinfo
  {note} {optical and Electrical Propertias of Glasses}\BibitemShut {NoStop}%
\bibitem [{\citenamefont {Hosono}\ \emph {et~al.}(1996)\citenamefont {Hosono},
  \citenamefont {Kikuchi}, \citenamefont {Ueda},\ and\ \citenamefont
  {Kawazoe}}]{12}%
  \BibitemOpen
  \bibfield  {author} {\bibinfo {author} {\bibfnamefont {H.}~\bibnamefont
  {Hosono}}, \bibinfo {author} {\bibfnamefont {N.}~\bibnamefont {Kikuchi}},
  \bibinfo {author} {\bibfnamefont {N.}~\bibnamefont {Ueda}}, \ and\ \bibinfo
  {author} {\bibfnamefont {H.}~\bibnamefont {Kawazoe}},\ }\bibfield  {title}
  {\enquote {\bibinfo {title} {{Working hypothesis to explore novel wide band
  gap electrically conducting amorphous oxides and examples}},}\ }\href
  {\doibase https://doi.org/10.1016/0022-3093(96)80019-6} {\bibfield  {journal}
  {\bibinfo  {journal} {Journal of Non-Crystalline Solids}\ }\textbf {\bibinfo
  {volume} {198-200}},\ \bibinfo {pages} {165 -- 169} (\bibinfo {year}
  {1996})},\ \bibinfo {note} {proceedings of the Sixteenth International
  Conference on Amorphous Semiconductors - Science and Technology}\BibitemShut
  {NoStop}%
\bibitem [{\citenamefont {Nomura}\ \emph {et~al.}(2004)\citenamefont {Nomura},
  \citenamefont {Ohta}, \citenamefont {Takagi}, \citenamefont {Kamiya},
  \citenamefont {Hirano},\ and\ \citenamefont {Hosono}}]{13}%
  \BibitemOpen
  \bibfield  {author} {\bibinfo {author} {\bibfnamefont {K.}~\bibnamefont
  {Nomura}}, \bibinfo {author} {\bibfnamefont {H.}~\bibnamefont {Ohta}},
  \bibinfo {author} {\bibfnamefont {A.}~\bibnamefont {Takagi}}, \bibinfo
  {author} {\bibfnamefont {T.}~\bibnamefont {Kamiya}}, \bibinfo {author}
  {\bibfnamefont {M.}~\bibnamefont {Hirano}}, \ and\ \bibinfo {author}
  {\bibfnamefont {H.}~\bibnamefont {Hosono}},\ }\bibfield  {title} {\enquote
  {\bibinfo {title} {{Room-Temperature Fabrication of Transparent Flexible
  Thin-Film Transistors Using Amorphous Oxide Semiconductors}},}\ }\href
  {\doibase 10.1038/nature03090} {\bibfield  {journal} {\bibinfo  {journal}
  {Nature}\ }\textbf {\bibinfo {volume} {432}},\ \bibinfo {pages} {488--92}
  (\bibinfo {year} {2004})}\BibitemShut {NoStop}%
\bibitem [{\citenamefont {Kamiya}, \citenamefont {Nomura},\ and\ \citenamefont
  {Hosono}(2009)}]{14}%
  \BibitemOpen
  \bibfield  {author} {\bibinfo {author} {\bibfnamefont {T.}~\bibnamefont
  {Kamiya}}, \bibinfo {author} {\bibfnamefont {K.}~\bibnamefont {Nomura}}, \
  and\ \bibinfo {author} {\bibfnamefont {H.}~\bibnamefont {Hosono}},\
  }\bibfield  {title} {\enquote {\bibinfo {title} {{Origins of High Mobility
  and Low Operation Voltage of Amorphous Oxide TFTs: Electronic Structure,
  Electron Transport, Defects and Doping}},}\ }\href {\doibase
  10.1109/JDT.2009.2021582} {\bibfield  {journal} {\bibinfo  {journal} {Journal
  of Display Technology}\ }\textbf {\bibinfo {volume} {5}},\ \bibinfo {pages}
  {273--288} (\bibinfo {year} {2009})}\BibitemShut {NoStop}%
\bibitem [{\citenamefont {Nomura}\ \emph
  {et~al.}(2007{\natexlab{a}})\citenamefont {Nomura}, \citenamefont {Kamiya},
  \citenamefont {Ohta}, \citenamefont {Uruga}, \citenamefont {Hirano},\ and\
  \citenamefont {Hosono}}]{15}%
  \BibitemOpen
  \bibfield  {author} {\bibinfo {author} {\bibfnamefont {K.}~\bibnamefont
  {Nomura}}, \bibinfo {author} {\bibfnamefont {T.}~\bibnamefont {Kamiya}},
  \bibinfo {author} {\bibfnamefont {H.}~\bibnamefont {Ohta}}, \bibinfo {author}
  {\bibfnamefont {T.}~\bibnamefont {Uruga}}, \bibinfo {author} {\bibfnamefont
  {M.}~\bibnamefont {Hirano}}, \ and\ \bibinfo {author} {\bibfnamefont
  {H.}~\bibnamefont {Hosono}},\ }\bibfield  {title} {\enquote {\bibinfo {title}
  {{Local coordination structure and electronic structure of the large electron
  mobility amorphous oxide semiconductor In-Ga-Zn-O: Experiment and ab initio
  calculations}},}\ }\href {\doibase 10.1103/PhysRevB.75.035212} {\bibfield
  {journal} {\bibinfo  {journal} {Phys. Rev. B}\ }\textbf {\bibinfo {volume}
  {75}},\ \bibinfo {pages} {035212} (\bibinfo {year}
  {2007}{\natexlab{a}})}\BibitemShut {NoStop}%
\bibitem [{\citenamefont {{Takagi}}\ \emph {et~al.}(2005)\citenamefont
  {{Takagi}}, \citenamefont {{Nomura}}, \citenamefont {{Ohta}}, \citenamefont
  {{Yanagi}}, \citenamefont {{Kamiya}}, \citenamefont {{Hirano}},\ and\
  \citenamefont {{Hosono}}}]{16}%
  \BibitemOpen
  \bibfield  {author} {\bibinfo {author} {\bibfnamefont {A.}~\bibnamefont
  {{Takagi}}}, \bibinfo {author} {\bibfnamefont {K.}~\bibnamefont {{Nomura}}},
  \bibinfo {author} {\bibfnamefont {H.}~\bibnamefont {{Ohta}}}, \bibinfo
  {author} {\bibfnamefont {H.}~\bibnamefont {{Yanagi}}}, \bibinfo {author}
  {\bibfnamefont {T.}~\bibnamefont {{Kamiya}}}, \bibinfo {author}
  {\bibfnamefont {M.}~\bibnamefont {{Hirano}}}, \ and\ \bibinfo {author}
  {\bibfnamefont {H.}~\bibnamefont {{Hosono}}},\ }\bibfield  {title} {\enquote
  {\bibinfo {title} {{Carrier transport and electronic structure in amorphous
  oxide semiconductor, a-InGaZnO}},}\ }\href {\doibase
  10.1016/j.tsf.2004.11.223} {\bibfield  {journal} {\bibinfo  {journal} {Thin
  Solid Films}\ }\textbf {\bibinfo {volume} {486}},\ \bibinfo {pages} {38--41}
  (\bibinfo {year} {2005})}\BibitemShut {NoStop}%
\bibitem [{\citenamefont {Yamaguchi}\ \emph {et~al.}(2009)\citenamefont
  {Yamaguchi}, \citenamefont {Taniguchi}, \citenamefont {Miyajima},\ and\
  \citenamefont {Ikeda}}]{17}%
  \BibitemOpen
  \bibfield  {author} {\bibinfo {author} {\bibfnamefont {N.}~\bibnamefont
  {Yamaguchi}}, \bibinfo {author} {\bibfnamefont {S.}~\bibnamefont
  {Taniguchi}}, \bibinfo {author} {\bibfnamefont {T.}~\bibnamefont {Miyajima}},
  \ and\ \bibinfo {author} {\bibfnamefont {M.}~\bibnamefont {Ikeda}},\
  }\bibfield  {title} {\enquote {\bibinfo {title} {{Optical and electrical
  properties of amorphous InGaZnO}},}\ }\href {\doibase 10.1116/1.3110022}
  {\bibfield  {journal} {\bibinfo  {journal} {Journal of Vacuum Science \&
  Technology B: Microelectronics and Nanometer Structures Processing,
  Measurement, and Phenomena}\ }\textbf {\bibinfo {volume} {27}},\ \bibinfo
  {pages} {1746--1748} (\bibinfo {year} {2009})}\BibitemShut {NoStop}%
\bibitem [{\citenamefont {Jeong}\ \emph {et~al.}(2008)\citenamefont {Jeong},
  \citenamefont {Jeong}, \citenamefont {Choi}, \citenamefont {Im},
  \citenamefont {Kim}, \citenamefont {Yang}, \citenamefont {Kang},
  \citenamefont {Kim}, \citenamefont {Ahn}, \citenamefont {Chung},
  \citenamefont {Kim}, \citenamefont {Gu}, \citenamefont {Park}, \citenamefont
  {Mo}, \citenamefont {Kim},\ and\ \citenamefont {Chung}}]{18}%
  \BibitemOpen
  \bibfield  {author} {\bibinfo {author} {\bibfnamefont {J.~K.}\ \bibnamefont
  {Jeong}}, \bibinfo {author} {\bibfnamefont {J.~H.}\ \bibnamefont {Jeong}},
  \bibinfo {author} {\bibfnamefont {J.~H.}\ \bibnamefont {Choi}}, \bibinfo
  {author} {\bibfnamefont {J.~S.}\ \bibnamefont {Im}}, \bibinfo {author}
  {\bibfnamefont {S.~H.}\ \bibnamefont {Kim}}, \bibinfo {author} {\bibfnamefont
  {H.~W.}\ \bibnamefont {Yang}}, \bibinfo {author} {\bibfnamefont {K.~N.}\
  \bibnamefont {Kang}}, \bibinfo {author} {\bibfnamefont {K.~S.}\ \bibnamefont
  {Kim}}, \bibinfo {author} {\bibfnamefont {T.~K.}\ \bibnamefont {Ahn}},
  \bibinfo {author} {\bibfnamefont {H.-J.}\ \bibnamefont {Chung}}, \bibinfo
  {author} {\bibfnamefont {M.}~\bibnamefont {Kim}}, \bibinfo {author}
  {\bibfnamefont {B.~S.}\ \bibnamefont {Gu}}, \bibinfo {author} {\bibfnamefont
  {J.-S.}\ \bibnamefont {Park}}, \bibinfo {author} {\bibfnamefont {Y.-G.}\
  \bibnamefont {Mo}}, \bibinfo {author} {\bibfnamefont {H.~D.}\ \bibnamefont
  {Kim}}, \ and\ \bibinfo {author} {\bibfnamefont {H.~K.}\ \bibnamefont
  {Chung}},\ }\bibfield  {title} {\enquote {\bibinfo {title} {{3.1:
  Distinguished Paper: 12.1-Inch WXGA AMOLED Display Driven by
  Indium-Gallium-Zinc Oxide TFTs Array}},}\ }\href {\doibase 10.1889/1.3069591}
  {\bibfield  {journal} {\bibinfo  {journal} {SID Symposium Digest of Technical
  Papers}\ }\textbf {\bibinfo {volume} {39}},\ \bibinfo {pages} {1--4}
  (\bibinfo {year} {2008})}\BibitemShut {NoStop}%
\bibitem [{\citenamefont {Ohara}\ \emph {et~al.}(2009)\citenamefont {Ohara},
  \citenamefont {Sasaki}, \citenamefont {Noda}, \citenamefont {Ito},
  \citenamefont {Sasaki}, \citenamefont {Toyosumi}, \citenamefont {Endo},
  \citenamefont {Yoshitomi}, \citenamefont {Sakata}, \citenamefont {Serikawa},\
  and\ \citenamefont {Yamazaki}}]{19}%
  \BibitemOpen
  \bibfield  {author} {\bibinfo {author} {\bibfnamefont {H.}~\bibnamefont
  {Ohara}}, \bibinfo {author} {\bibfnamefont {T.}~\bibnamefont {Sasaki}},
  \bibinfo {author} {\bibfnamefont {K.}~\bibnamefont {Noda}}, \bibinfo {author}
  {\bibfnamefont {S.}~\bibnamefont {Ito}}, \bibinfo {author} {\bibfnamefont
  {M.}~\bibnamefont {Sasaki}}, \bibinfo {author} {\bibfnamefont
  {Y.}~\bibnamefont {Toyosumi}}, \bibinfo {author} {\bibfnamefont
  {Y.}~\bibnamefont {Endo}}, \bibinfo {author} {\bibfnamefont {S.}~\bibnamefont
  {Yoshitomi}}, \bibinfo {author} {\bibfnamefont {J.}~\bibnamefont {Sakata}},
  \bibinfo {author} {\bibfnamefont {T.}~\bibnamefont {Serikawa}}, \ and\
  \bibinfo {author} {\bibfnamefont {S.}~\bibnamefont {Yamazaki}},\ }\bibfield
  {title} {\enquote {\bibinfo {title} {{21.3: 4.0 In. QVGA AMOLED Display Using
  In-Ga-Zn-Oxide TFTs with a Novel Passivation Layer}},}\ }\href {\doibase
  10.1889/1.3256764} {\bibfield  {journal} {\bibinfo  {journal} {SID Symposium
  Digest of Technical Papers}\ }\textbf {\bibinfo {volume} {40}},\ \bibinfo
  {pages} {284--287} (\bibinfo {year} {2009})}\BibitemShut {NoStop}%
\bibitem [{\citenamefont {Lee}\ \emph {et~al.}(2011)\citenamefont {Lee},
  \citenamefont {Nathan}, \citenamefont {Robertson}, \citenamefont
  {Ghaffarzadeh}, \citenamefont {Pepper}, \citenamefont {Jeon}, \citenamefont
  {Kim}, \citenamefont {Song}, \citenamefont {Chung},\ and\ \citenamefont
  {Kim}}]{20}%
  \BibitemOpen
  \bibfield  {author} {\bibinfo {author} {\bibfnamefont {S.}~\bibnamefont
  {Lee}}, \bibinfo {author} {\bibfnamefont {A.}~\bibnamefont {Nathan}},
  \bibinfo {author} {\bibfnamefont {J.}~\bibnamefont {Robertson}}, \bibinfo
  {author} {\bibfnamefont {K.}~\bibnamefont {Ghaffarzadeh}}, \bibinfo {author}
  {\bibfnamefont {M.}~\bibnamefont {Pepper}}, \bibinfo {author} {\bibfnamefont
  {S.}~\bibnamefont {Jeon}}, \bibinfo {author} {\bibfnamefont {C.}~\bibnamefont
  {Kim}}, \bibinfo {author} {\bibfnamefont {I.-H.}\ \bibnamefont {Song}},
  \bibinfo {author} {\bibfnamefont {U.-I.}\ \bibnamefont {Chung}}, \ and\
  \bibinfo {author} {\bibfnamefont {K.}~\bibnamefont {Kim}},\ }\bibfield
  {title} {\enquote {\bibinfo {title} {{Temperature Dependent Carrier Transport
  in Amorphous Oxide Field Effect Transistors }},}\ }\href
  {http://discovery.ucl.ac.uk/id/eprint/1400552} {\bibfield  {journal}
  {\bibinfo  {journal} {IEEE International Electron Device Meeting (IEDM) Tech.
  Dig.}\ } (\bibinfo {year} {2011})}\BibitemShut {NoStop}%
\bibitem [{\citenamefont {Jeon}\ \emph {et~al.}(2012)\citenamefont {Jeon},
  \citenamefont {Ahn}, \citenamefont {Song}, \citenamefont {Kim}, \citenamefont
  {Chung}, \citenamefont {Lee}, \citenamefont {Yoo}, \citenamefont {Nathan},
  \citenamefont {Lee}, \citenamefont {Ghaffarzadeh}, \citenamefont
  {Robertson},\ and\ \citenamefont {Kim}}]{21}%
  \BibitemOpen
  \bibfield  {author} {\bibinfo {author} {\bibfnamefont {S.}~\bibnamefont
  {Jeon}}, \bibinfo {author} {\bibfnamefont {S.-E.}\ \bibnamefont {Ahn}},
  \bibinfo {author} {\bibfnamefont {I.}~\bibnamefont {Song}}, \bibinfo {author}
  {\bibfnamefont {C.~J.}\ \bibnamefont {Kim}}, \bibinfo {author} {\bibfnamefont
  {U.-I.}\ \bibnamefont {Chung}}, \bibinfo {author} {\bibfnamefont
  {E.}~\bibnamefont {Lee}}, \bibinfo {author} {\bibfnamefont {I.}~\bibnamefont
  {Yoo}}, \bibinfo {author} {\bibfnamefont {A.}~\bibnamefont {Nathan}},
  \bibinfo {author} {\bibfnamefont {S.}~\bibnamefont {Lee}}, \bibinfo {author}
  {\bibfnamefont {K.}~\bibnamefont {Ghaffarzadeh}}, \bibinfo {author}
  {\bibfnamefont {J.}~\bibnamefont {Robertson}}, \ and\ \bibinfo {author}
  {\bibfnamefont {K.}~\bibnamefont {Kim}},\ }\bibfield  {title} {\enquote
  {\bibinfo {title} {{Gated three-terminal device architecture to eliminate
  persistent photoconductivity in oxide semiconductor photosensor arrays}},}\
  }\href {\doibase 10.1038/nmat3256} {\bibfield  {journal} {\bibinfo  {journal}
  {Nature Materials}\ }\textbf {\bibinfo {volume} {11}},\ \bibinfo {pages}
  {301--305} (\bibinfo {year} {2012})}\BibitemShut {NoStop}%
\bibitem [{\citenamefont {Münzenrieder}\ \emph {et~al.}(2013)\citenamefont
  {Münzenrieder}, \citenamefont {Petti}, \citenamefont {Zysset}, \citenamefont
  {Kinkeldei}, \citenamefont {Salvatore},\ and\ \citenamefont {Tröster}}]{22}%
  \BibitemOpen
  \bibfield  {author} {\bibinfo {author} {\bibfnamefont {N.}~\bibnamefont
  {Münzenrieder}}, \bibinfo {author} {\bibfnamefont {L.}~\bibnamefont
  {Petti}}, \bibinfo {author} {\bibfnamefont {C.}~\bibnamefont {Zysset}},
  \bibinfo {author} {\bibfnamefont {T.}~\bibnamefont {Kinkeldei}}, \bibinfo
  {author} {\bibfnamefont {G.~A.}\ \bibnamefont {Salvatore}}, \ and\ \bibinfo
  {author} {\bibfnamefont {G.}~\bibnamefont {Tröster}},\ }\bibfield  {title}
  {\enquote {\bibinfo {title} {{Flexible Self-Aligned Amorphous InGaZnO
  Thin-Film Transistors With Submicrometer Channel Length and a Transit
  Frequency of 135 MHz}},}\ }\href {\doibase 10.1109/TED.2013.2274575}
  {\bibfield  {journal} {\bibinfo  {journal} {IEEE Transactions on Electron
  Devices}\ }\textbf {\bibinfo {volume} {60}},\ \bibinfo {pages} {2815--2820}
  (\bibinfo {year} {2013})}\BibitemShut {NoStop}%
\bibitem [{\citenamefont {Fortunato}, \citenamefont {Barquinha},\ and\
  \citenamefont {Martins}(2012)}]{23}%
  \BibitemOpen
  \bibfield  {author} {\bibinfo {author} {\bibfnamefont {E.}~\bibnamefont
  {Fortunato}}, \bibinfo {author} {\bibfnamefont {P.}~\bibnamefont
  {Barquinha}}, \ and\ \bibinfo {author} {\bibfnamefont {R.}~\bibnamefont
  {Martins}},\ }\bibfield  {title} {\enquote {\bibinfo {title} {{Oxide
  Semiconductor Thin-Film Transistors: A Review of Recent Advances}},}\ }\href
  {\doibase 10.1002/adma.201103228} {\bibfield  {journal} {\bibinfo  {journal}
  {Advanced Materials}\ }\textbf {\bibinfo {volume} {24}},\ \bibinfo {pages}
  {2945--2986} (\bibinfo {year} {2012})}\BibitemShut {NoStop}%
\bibitem [{\citenamefont {Ahn}\ \emph {et~al.}(2012{\natexlab{a}})\citenamefont
  {Ahn}, \citenamefont {Jeon}, \citenamefont {Song}, \citenamefont {Jeon},
  \citenamefont {Kim}, \citenamefont {Kim}, \citenamefont {Lim}, \citenamefont
  {Jeong}, \citenamefont {Goh}, \citenamefont {Yeon}, \citenamefont {Lee},
  \citenamefont {Kim}, \citenamefont {Lee}, \citenamefont {Song}, \citenamefont
  {Nathan}, \citenamefont {Lee},\ and\ \citenamefont {Chung}}]{24}%
  \BibitemOpen
  \bibfield  {author} {\bibinfo {author} {\bibfnamefont {S.-e.}\ \bibnamefont
  {Ahn}}, \bibinfo {author} {\bibfnamefont {S.}~\bibnamefont {Jeon}}, \bibinfo
  {author} {\bibfnamefont {I.}~\bibnamefont {Song}}, \bibinfo {author}
  {\bibfnamefont {Y.}~\bibnamefont {Jeon}}, \bibinfo {author} {\bibfnamefont
  {Y.}~\bibnamefont {Kim}}, \bibinfo {author} {\bibfnamefont {C.}~\bibnamefont
  {Kim}}, \bibinfo {author} {\bibfnamefont {J.}~\bibnamefont {Lim}}, \bibinfo
  {author} {\bibfnamefont {W.}~\bibnamefont {Jeong}}, \bibinfo {author}
  {\bibfnamefont {J.}~\bibnamefont {Goh}}, \bibinfo {author} {\bibfnamefont
  {S.}~\bibnamefont {Yeon}}, \bibinfo {author} {\bibfnamefont {C.}~\bibnamefont
  {Lee}}, \bibinfo {author} {\bibfnamefont {J.-h.}\ \bibnamefont {Kim}},
  \bibinfo {author} {\bibfnamefont {J.-h.}\ \bibnamefont {Lee}}, \bibinfo
  {author} {\bibfnamefont {J.}~\bibnamefont {Song}}, \bibinfo {author}
  {\bibfnamefont {A.}~\bibnamefont {Nathan}}, \bibinfo {author} {\bibfnamefont
  {S.}~\bibnamefont {Lee}}, \ and\ \bibinfo {author} {\bibfnamefont {U.-i.}\
  \bibnamefont {Chung}},\ }\bibfield  {title} {\enquote {\bibinfo {title}
  {{25.2: Photo-Sensor Thin Film Transistor based on Double Metal-Oxide Layer
  for In-cell Remote Touch Screen}},}\ }\href {\doibase
  10.1002/j.2168-0159.2012.tb05783.x} {\bibfield  {journal} {\bibinfo
  {journal} {SID Symposium Digest of Technical Papers}\ }\textbf {\bibinfo
  {volume} {43}},\ \bibinfo {pages} {334--337} (\bibinfo {year}
  {2012}{\natexlab{a}})}\BibitemShut {NoStop}%
\bibitem [{\citenamefont {Ahn}\ \emph {et~al.}(2012{\natexlab{b}})\citenamefont
  {Ahn}, \citenamefont {Song}, \citenamefont {Jeon}, \citenamefont {Jeon},
  \citenamefont {Kim}, \citenamefont {Kim}, \citenamefont {Ryu}, \citenamefont
  {Lee}, \citenamefont {Nathan}, \citenamefont {Lee}, \citenamefont {Kim},\
  and\ \citenamefont {Chung}}]{25}%
  \BibitemOpen
  \bibfield  {author} {\bibinfo {author} {\bibfnamefont {S.-E.}\ \bibnamefont
  {Ahn}}, \bibinfo {author} {\bibfnamefont {I.}~\bibnamefont {Song}}, \bibinfo
  {author} {\bibfnamefont {S.}~\bibnamefont {Jeon}}, \bibinfo {author}
  {\bibfnamefont {Y.~W.}\ \bibnamefont {Jeon}}, \bibinfo {author}
  {\bibfnamefont {Y.}~\bibnamefont {Kim}}, \bibinfo {author} {\bibfnamefont
  {C.}~\bibnamefont {Kim}}, \bibinfo {author} {\bibfnamefont {B.}~\bibnamefont
  {Ryu}}, \bibinfo {author} {\bibfnamefont {J.-H.}\ \bibnamefont {Lee}},
  \bibinfo {author} {\bibfnamefont {A.}~\bibnamefont {Nathan}}, \bibinfo
  {author} {\bibfnamefont {S.}~\bibnamefont {Lee}}, \bibinfo {author}
  {\bibfnamefont {G.~T.}\ \bibnamefont {Kim}}, \ and\ \bibinfo {author}
  {\bibfnamefont {U.-I.}\ \bibnamefont {Chung}},\ }\bibfield  {title} {\enquote
  {\bibinfo {title} {{Metal Oxide Thin Film Phototransistor for Remote Touch
  Interactive Displays}},}\ }\href {\doibase 10.1002/adma.201200293} {\bibfield
   {journal} {\bibinfo  {journal} {Advanced Materials}\ }\textbf {\bibinfo
  {volume} {24}},\ \bibinfo {pages} {2631--2636} (\bibinfo {year}
  {2012}{\natexlab{b}})}\BibitemShut {NoStop}%
\bibitem [{\citenamefont {Ye}\ \emph {et~al.}(2013)\citenamefont {Ye},
  \citenamefont {Lim}, \citenamefont {You}, \citenamefont {Scheer},
  \citenamefont {Gaur}, \citenamefont {Hsu}, \citenamefont {Liu}, \citenamefont
  {Yim}, \citenamefont {Hosokawa},\ and\ \citenamefont {White}}]{26}%
  \BibitemOpen
  \bibfield  {author} {\bibinfo {author} {\bibfnamefont {Y.}~\bibnamefont
  {Ye}}, \bibinfo {author} {\bibfnamefont {R.}~\bibnamefont {Lim}}, \bibinfo
  {author} {\bibfnamefont {H.}~\bibnamefont {You}}, \bibinfo {author}
  {\bibfnamefont {E.}~\bibnamefont {Scheer}}, \bibinfo {author} {\bibfnamefont
  {A.}~\bibnamefont {Gaur}}, \bibinfo {author} {\bibfnamefont {H.-c.}\
  \bibnamefont {Hsu}}, \bibinfo {author} {\bibfnamefont {J.}~\bibnamefont
  {Liu}}, \bibinfo {author} {\bibfnamefont {D.~K.}\ \bibnamefont {Yim}},
  \bibinfo {author} {\bibfnamefont {A.}~\bibnamefont {Hosokawa}}, \ and\
  \bibinfo {author} {\bibfnamefont {J.~M.}\ \bibnamefont {White}},\ }\bibfield
  {title} {\enquote {\bibinfo {title} {{4.2: Invited Paper: Development of High
  Mobility Zinc Oxynitride Thin Film Transistors}},}\ }\href {\doibase
  10.1002/j.2168-0159.2013.tb06127.x} {\bibfield  {journal} {\bibinfo
  {journal} {SID Symposium Digest of Technical Papers}\ }\textbf {\bibinfo
  {volume} {44}},\ \bibinfo {pages} {14--17} (\bibinfo {year}
  {2013})}\BibitemShut {NoStop}%
\bibitem [{\citenamefont {Park}\ \emph {et~al.}(2016)\citenamefont {Park},
  \citenamefont {Soo~Kim}, \citenamefont {Ok}, \citenamefont {Chang~Park},
  \citenamefont {Kim}, \citenamefont {Park},\ and\ \citenamefont {Kim}}]{27}%
  \BibitemOpen
  \bibfield  {author} {\bibinfo {author} {\bibfnamefont {J.}~\bibnamefont
  {Park}}, \bibinfo {author} {\bibfnamefont {Y.}~\bibnamefont {Soo~Kim}},
  \bibinfo {author} {\bibfnamefont {K.-C.}\ \bibnamefont {Ok}}, \bibinfo
  {author} {\bibfnamefont {Y.}~\bibnamefont {Chang~Park}}, \bibinfo {author}
  {\bibfnamefont {H.}~\bibnamefont {Kim}}, \bibinfo {author} {\bibfnamefont
  {J.-S.}\ \bibnamefont {Park}}, \ and\ \bibinfo {author} {\bibfnamefont
  {H.-S.}\ \bibnamefont {Kim}},\ }\bibfield  {title} {\enquote {\bibinfo
  {title} {{A study on the electron transport properties of ZnON semiconductors
  with respect to the relative anion content}},}\ }\href {\doibase
  10.1038/srep24787} {\bibfield  {journal} {\bibinfo  {journal} {Scientific
  Reports}\ }\textbf {\bibinfo {volume} {6}},\ \bibinfo {pages} {24787}
  (\bibinfo {year} {2016})}\BibitemShut {NoStop}%
\bibitem [{\citenamefont {Ye}, \citenamefont {Lim},\ and\ \citenamefont
  {White}(2009)}]{28}%
  \BibitemOpen
  \bibfield  {author} {\bibinfo {author} {\bibfnamefont {Y.}~\bibnamefont
  {Ye}}, \bibinfo {author} {\bibfnamefont {R.}~\bibnamefont {Lim}}, \ and\
  \bibinfo {author} {\bibfnamefont {J.~M.}\ \bibnamefont {White}},\ }\bibfield
  {title} {\enquote {\bibinfo {title} {{High mobility amorphous zinc oxynitride
  semiconductor material for thin film transistors}},}\ }\href {\doibase
  10.1063/1.3236663} {\bibfield  {journal} {\bibinfo  {journal} {Journal of
  Applied Physics}\ }\textbf {\bibinfo {volume} {106}},\ \bibinfo {pages}
  {074512} (\bibinfo {year} {2009})}\BibitemShut {NoStop}%
\bibitem [{\citenamefont {Kim}\ \emph {et~al.}(2013)\citenamefont {Kim},
  \citenamefont {Ho~Jeon}, \citenamefont {Park}, \citenamefont {Sang~Kim},
  \citenamefont {Seok~Son}, \citenamefont {Seon}, \citenamefont {Seo},
  \citenamefont {Kim}, \citenamefont {Lee}, \citenamefont {Chung},
  \citenamefont {Lee}, \citenamefont {Han}, \citenamefont {Ryu}, \citenamefont
  {Lee},\ and\ \citenamefont {Kim}}]{29}%
  \BibitemOpen
  \bibfield  {author} {\bibinfo {author} {\bibfnamefont {H.-S.}\ \bibnamefont
  {Kim}}, \bibinfo {author} {\bibfnamefont {S.}~\bibnamefont {Ho~Jeon}},
  \bibinfo {author} {\bibfnamefont {J.}~\bibnamefont {Park}}, \bibinfo {author}
  {\bibfnamefont {T.}~\bibnamefont {Sang~Kim}}, \bibinfo {author}
  {\bibfnamefont {K.}~\bibnamefont {Seok~Son}}, \bibinfo {author}
  {\bibfnamefont {J.-B.}\ \bibnamefont {Seon}}, \bibinfo {author}
  {\bibfnamefont {S.-J.}\ \bibnamefont {Seo}}, \bibinfo {author} {\bibfnamefont
  {S.-J.}\ \bibnamefont {Kim}}, \bibinfo {author} {\bibfnamefont
  {E.}~\bibnamefont {Lee}}, \bibinfo {author} {\bibfnamefont {J.}~\bibnamefont
  {Chung}}, \bibinfo {author} {\bibfnamefont {H.}~\bibnamefont {Lee}}, \bibinfo
  {author} {\bibfnamefont {S.}~\bibnamefont {Han}}, \bibinfo {author}
  {\bibfnamefont {M.}~\bibnamefont {Ryu}}, \bibinfo {author} {\bibfnamefont
  {S.}~\bibnamefont {Lee}}, \ and\ \bibinfo {author} {\bibfnamefont
  {K.}~\bibnamefont {Kim}},\ }\bibfield  {title} {\enquote {\bibinfo {title}
  {{Anion control as a strategy to achieve high-mobility and high-stability
  oxide thin-film transistors}},}\ }\href {\doibase 10.1038/srep01459}
  {\bibfield  {journal} {\bibinfo  {journal} {Scientific reports}\ }\textbf
  {\bibinfo {volume} {3}},\ \bibinfo {pages} {1459} (\bibinfo {year}
  {2013})}\BibitemShut {NoStop}%
\bibitem [{\citenamefont {Lee}\ \emph {et~al.}(2015{\natexlab{a}})\citenamefont
  {Lee}, \citenamefont {Kim}, \citenamefont {Benayad}, \citenamefont {Kim},
  \citenamefont {Jeon},\ and\ \citenamefont {Park}}]{30}%
  \BibitemOpen
  \bibfield  {author} {\bibinfo {author} {\bibfnamefont {E.}~\bibnamefont
  {Lee}}, \bibinfo {author} {\bibfnamefont {T.}~\bibnamefont {Kim}}, \bibinfo
  {author} {\bibfnamefont {A.}~\bibnamefont {Benayad}}, \bibinfo {author}
  {\bibfnamefont {H.}~\bibnamefont {Kim}}, \bibinfo {author} {\bibfnamefont
  {S.}~\bibnamefont {Jeon}}, \ and\ \bibinfo {author} {\bibfnamefont {G.-S.}\
  \bibnamefont {Park}},\ }\bibfield  {title} {\enquote {\bibinfo {title} {{Ar
  plasma treated ZnON transistor for future thin film electronics}},}\ }\href
  {\doibase 10.1063/1.4930827} {\bibfield  {journal} {\bibinfo  {journal}
  {Applied Physics Letters}\ }\textbf {\bibinfo {volume} {107}},\ \bibinfo
  {pages} {122105} (\bibinfo {year} {2015}{\natexlab{a}})}\BibitemShut
  {NoStop}%
\bibitem [{\citenamefont {Lee}\ \emph {et~al.}(2016)\citenamefont {Lee},
  \citenamefont {Kim}, \citenamefont {Benayad}, \citenamefont {Hur},
  \citenamefont {Park},\ and\ \citenamefont {Jeon}}]{31}%
  \BibitemOpen
  \bibfield  {author} {\bibinfo {author} {\bibfnamefont {E.}~\bibnamefont
  {Lee}}, \bibinfo {author} {\bibfnamefont {T.}~\bibnamefont {Kim}}, \bibinfo
  {author} {\bibfnamefont {A.}~\bibnamefont {Benayad}}, \bibinfo {author}
  {\bibfnamefont {J.}~\bibnamefont {Hur}}, \bibinfo {author} {\bibfnamefont
  {G.-S.}\ \bibnamefont {Park}}, \ and\ \bibinfo {author} {\bibfnamefont
  {S.}~\bibnamefont {Jeon}},\ }\bibfield  {title} {\enquote {\bibinfo {title}
  {{High mobility and high stability glassy metal-oxynitride materials and
  devices}},}\ }\href {\doibase 10.1038/srep23940} {\bibfield  {journal}
  {\bibinfo  {journal} {Scientific Reports}\ }\textbf {\bibinfo {volume} {6}},\
  \bibinfo {pages} {23940} (\bibinfo {year} {2016})}\BibitemShut {NoStop}%
\bibitem [{\citenamefont {Lee}\ \emph {et~al.}(2015{\natexlab{b}})\citenamefont
  {Lee}, \citenamefont {Nathan}, \citenamefont {Ye}, \citenamefont {Guo},\ and\
  \citenamefont {Robertson}}]{32}%
  \BibitemOpen
  \bibfield  {author} {\bibinfo {author} {\bibfnamefont {S.}~\bibnamefont
  {Lee}}, \bibinfo {author} {\bibfnamefont {A.}~\bibnamefont {Nathan}},
  \bibinfo {author} {\bibfnamefont {Y.}~\bibnamefont {Ye}}, \bibinfo {author}
  {\bibfnamefont {Y.}~\bibnamefont {Guo}}, \ and\ \bibinfo {author}
  {\bibfnamefont {J.}~\bibnamefont {Robertson}},\ }\bibfield  {title} {\enquote
  {\bibinfo {title} {{Localized Tail States and Electron Mobility in Amorphous
  ZnON Thin Film Transistors}},}\ }\href {\doibase 10.1038/srep13467}
  {\bibfield  {journal} {\bibinfo  {journal} {Scientific reports}\ }\textbf
  {\bibinfo {volume} {5}},\ \bibinfo {pages} {13467} (\bibinfo {year}
  {2015}{\natexlab{b}})}\BibitemShut {NoStop}%
\bibitem [{\citenamefont {Yamazaki}\ \emph {et~al.}(2016)\citenamefont
  {Yamazaki}, \citenamefont {Shigematsu}, \citenamefont {Hirose}, \citenamefont
  {Nakao}, \citenamefont {Harayama}, \citenamefont {Sekiba},\ and\
  \citenamefont {Hasegawa}}]{33}%
  \BibitemOpen
  \bibfield  {author} {\bibinfo {author} {\bibfnamefont {T.}~\bibnamefont
  {Yamazaki}}, \bibinfo {author} {\bibfnamefont {K.}~\bibnamefont
  {Shigematsu}}, \bibinfo {author} {\bibfnamefont {Y.}~\bibnamefont {Hirose}},
  \bibinfo {author} {\bibfnamefont {S.}~\bibnamefont {Nakao}}, \bibinfo
  {author} {\bibfnamefont {I.}~\bibnamefont {Harayama}}, \bibinfo {author}
  {\bibfnamefont {D.}~\bibnamefont {Sekiba}}, \ and\ \bibinfo {author}
  {\bibfnamefont {T.}~\bibnamefont {Hasegawa}},\ }\bibfield  {title} {\enquote
  {\bibinfo {title} {{Amorphous ZnO$_x$N$_y$ thin films with high electron Hall
  mobility exceeding 200 cm$^{2+}$V$^{-1}$s$^{-1}$}},}\ }\href {\doibase
  10.1063/1.4973203} {\bibfield  {journal} {\bibinfo  {journal} {Applied
  Physics Letters}\ }\textbf {\bibinfo {volume} {109}},\ \bibinfo {pages}
  {262101} (\bibinfo {year} {2016})}\BibitemShut {NoStop}%
\bibitem [{\citenamefont {Glass}, \citenamefont {Oganov},\ and\ \citenamefont
  {Hansen}(2006)}]{34}%
  \BibitemOpen
  \bibfield  {author} {\bibinfo {author} {\bibfnamefont {C.~W.}\ \bibnamefont
  {Glass}}, \bibinfo {author} {\bibfnamefont {A.~R.}\ \bibnamefont {Oganov}}, \
  and\ \bibinfo {author} {\bibfnamefont {N.}~\bibnamefont {Hansen}},\
  }\bibfield  {title} {\enquote {\bibinfo {title} {{USPEX-Evolutionary crystal
  structure prediction", journal = "Computer Physics Communications}},}\ }\href
  {\doibase https://doi.org/10.1016/j.cpc.2006.07.020} {\ \textbf {\bibinfo
  {volume} {175}},\ \bibinfo {pages} {713 -- 720} (\bibinfo {year}
  {2006})}\BibitemShut {NoStop}%
\bibitem [{\citenamefont {Oganov}, \citenamefont {Lyakhov},\ and\ \citenamefont
  {Valle}(2011)}]{35}%
  \BibitemOpen
  \bibfield  {author} {\bibinfo {author} {\bibfnamefont {A.~R.}\ \bibnamefont
  {Oganov}}, \bibinfo {author} {\bibfnamefont {A.~O.}\ \bibnamefont {Lyakhov}},
  \ and\ \bibinfo {author} {\bibfnamefont {M.}~\bibnamefont {Valle}},\
  }\bibfield  {title} {\enquote {\bibinfo {title} {{How Evolutionary Crystal
  Structure Prediction Works—and Why}},}\ }\href {\doibase 10.1021/ar1001318}
  {\bibfield  {journal} {\bibinfo  {journal} {Accounts of Chemical Research}\
  }\textbf {\bibinfo {volume} {44}},\ \bibinfo {pages} {227--237} (\bibinfo
  {year} {2011})},\ \bibinfo {note} {pMID: 21361336}\BibitemShut {NoStop}%
\bibitem [{\citenamefont {Lyakhov}\ \emph {et~al.}(2013)\citenamefont
  {Lyakhov}, \citenamefont {Oganov}, \citenamefont {Stokes},\ and\
  \citenamefont {Zhu}}]{36}%
  \BibitemOpen
  \bibfield  {author} {\bibinfo {author} {\bibfnamefont {A.~O.}\ \bibnamefont
  {Lyakhov}}, \bibinfo {author} {\bibfnamefont {A.~R.}\ \bibnamefont {Oganov}},
  \bibinfo {author} {\bibfnamefont {H.~T.}\ \bibnamefont {Stokes}}, \ and\
  \bibinfo {author} {\bibfnamefont {Q.}~\bibnamefont {Zhu}},\ }\bibfield
  {title} {\enquote {\bibinfo {title} {{New developments in evolutionary
  structure prediction algorithm USPEX}},}\ }\href {\doibase
  https://doi.org/10.1016/j.cpc.2012.12.009} {\bibfield  {journal} {\bibinfo
  {journal} {Computer Physics Communications}\ }\textbf {\bibinfo {volume}
  {184}},\ \bibinfo {pages} {1172 -- 1182} (\bibinfo {year}
  {2013})}\BibitemShut {NoStop}%
\bibitem [{\citenamefont {Nahas}, \citenamefont {Gaur},\ and\ \citenamefont
  {Bhowmick}(2016)}]{37}%
  \BibitemOpen
  \bibfield  {author} {\bibinfo {author} {\bibfnamefont {S.}~\bibnamefont
  {Nahas}}, \bibinfo {author} {\bibfnamefont {A.}~\bibnamefont {Gaur}}, \ and\
  \bibinfo {author} {\bibfnamefont {S.}~\bibnamefont {Bhowmick}},\ }\bibfield
  {title} {\enquote {\bibinfo {title} {{First principles prediction of
  amorphous phases using evolutionary algorithms}},}\ }\href {\doibase
  10.1063/1.4955105} {\bibfield  {journal} {\bibinfo  {journal} {The Journal of
  Chemical Physics}\ }\textbf {\bibinfo {volume} {145}},\ \bibinfo {pages}
  {014106} (\bibinfo {year} {2016})}\BibitemShut {NoStop}%
\bibitem [{\citenamefont {Hanada}(2009)}]{42}%
  \BibitemOpen
  \bibfield  {author} {\bibinfo {author} {\bibfnamefont {T.}~\bibnamefont
  {Hanada}},\ }\bibfield  {title} {\enquote {\bibinfo {title} {{Basic
  Properties of {ZnO}, {GaN}, and {Related Materials}}},}\ }in\ \href@noop {}
  {\emph {\bibinfo {booktitle} {Oxide and Nitride Semiconductors}}}\ (\bibinfo
  {publisher} {Springer},\ \bibinfo {year} {2009})\ pp.\ \bibinfo {pages}
  {1--19}\BibitemShut {NoStop}%
\bibitem [{\citenamefont {Partin}, \citenamefont {Williams},\ and\
  \citenamefont {O'Keeffe}(1997)}]{44}%
  \BibitemOpen
  \bibfield  {author} {\bibinfo {author} {\bibfnamefont {D.}~\bibnamefont
  {Partin}}, \bibinfo {author} {\bibfnamefont {D.}~\bibnamefont {Williams}}, \
  and\ \bibinfo {author} {\bibfnamefont {M.}~\bibnamefont {O'Keeffe}},\
  }\bibfield  {title} {\enquote {\bibinfo {title} {{The crystal structures of
  Mg$_3$N$_2$ and Zn$_3$N$_2$}},}\ }\href {\doibase 10.1006/jssc.1997.7407}
  {\bibfield  {journal} {\bibinfo  {journal} {Journal of Solid State
  Chemistry}\ }\textbf {\bibinfo {volume} {132}},\ \bibinfo {pages} {56–59}
  (\bibinfo {year} {1997})}\BibitemShut {NoStop}%
\bibitem [{\citenamefont {Wang}(2004)}]{47}%
  \BibitemOpen
  \bibfield  {author} {\bibinfo {author} {\bibfnamefont {Z.}~\bibnamefont
  {Wang}},\ }\bibfield  {title} {\enquote {\bibinfo {title} {{Zinc oxide
  nanostructures: growth, properties and applications}},}\ }\href {\doibase
  10.1088/0953-8984/16/25/R01} {\bibfield  {journal} {\bibinfo  {journal} {J.
  Phys.: Condens. Matter}\ }\textbf {\bibinfo {volume} {16}},\ \bibinfo {pages}
  {829--858} (\bibinfo {year} {2004})}\BibitemShut {NoStop}%
\bibitem [{\citenamefont {Kohn}\ and\ \citenamefont {Sham}(1965)}]{38}%
  \BibitemOpen
  \bibfield  {author} {\bibinfo {author} {\bibfnamefont {W.}~\bibnamefont
  {Kohn}}\ and\ \bibinfo {author} {\bibfnamefont {L.~J.}\ \bibnamefont
  {Sham}},\ }\bibfield  {title} {\enquote {\bibinfo {title} {{Self-Consistent
  Equations Including Exchange and Correlation Effects}},}\ }\href {\doibase
  10.1103/PhysRev.140.A1133} {\bibfield  {journal} {\bibinfo  {journal} {Phys.
  Rev.}\ }\textbf {\bibinfo {volume} {140}},\ \bibinfo {pages} {A1133--A1138}
  (\bibinfo {year} {1965})}\BibitemShut {NoStop}%
\bibitem [{\citenamefont {Perdew}, \citenamefont {Burke},\ and\ \citenamefont
  {Ernzerhof}(1996)}]{39}%
  \BibitemOpen
  \bibfield  {author} {\bibinfo {author} {\bibfnamefont {J.~P.}\ \bibnamefont
  {Perdew}}, \bibinfo {author} {\bibfnamefont {K.}~\bibnamefont {Burke}}, \
  and\ \bibinfo {author} {\bibfnamefont {M.}~\bibnamefont {Ernzerhof}},\
  }\bibfield  {title} {\enquote {\bibinfo {title} {{Generalized Gradient
  Approximation Made Simple}},}\ }\href {\doibase 10.1103/PhysRevLett.77.3865}
  {\bibfield  {journal} {\bibinfo  {journal} {Phys. Rev. Lett.}\ }\textbf
  {\bibinfo {volume} {77}},\ \bibinfo {pages} {3865--3868} (\bibinfo {year}
  {1996})}\BibitemShut {NoStop}%
\bibitem [{\citenamefont {Giannozzi}\ \emph {et~al.}(2009)\citenamefont
  {Giannozzi}, \citenamefont {Baroni}, \citenamefont {Bonini}, \citenamefont
  {Calandra}, \citenamefont {Car}, \citenamefont {Cavazzoni}, \citenamefont
  {Ceresoli}, \citenamefont {Chiarotti}, \citenamefont {Cococcioni},
  \citenamefont {Dabo}, \citenamefont {Corso}, \citenamefont {de~Gironcoli},
  \citenamefont {Fabris}, \citenamefont {Fratesi}, \citenamefont {Gebauer},
  \citenamefont {Gerstmann}, \citenamefont {Gougoussis}, \citenamefont
  {Kokalj}, \citenamefont {Lazzeri}, \citenamefont {Martin-Samos},
  \citenamefont {Marzari}, \citenamefont {Mauri}, \citenamefont {Mazzarello},
  \citenamefont {Paolini}, \citenamefont {Pasquarello}, \citenamefont
  {Paulatto}, \citenamefont {Sbraccia}, \citenamefont {Scandolo}, \citenamefont
  {Sclauzero}, \citenamefont {Seitsonen}, \citenamefont {Smogunov},
  \citenamefont {Umari},\ and\ \citenamefont {Wentzcovitch}}]{40}%
  \BibitemOpen
  \bibfield  {author} {\bibinfo {author} {\bibfnamefont {P.}~\bibnamefont
  {Giannozzi}}, \bibinfo {author} {\bibfnamefont {S.}~\bibnamefont {Baroni}},
  \bibinfo {author} {\bibfnamefont {N.}~\bibnamefont {Bonini}}, \bibinfo
  {author} {\bibfnamefont {M.}~\bibnamefont {Calandra}}, \bibinfo {author}
  {\bibfnamefont {R.}~\bibnamefont {Car}}, \bibinfo {author} {\bibfnamefont
  {C.}~\bibnamefont {Cavazzoni}}, \bibinfo {author} {\bibfnamefont
  {D.}~\bibnamefont {Ceresoli}}, \bibinfo {author} {\bibfnamefont {G.~L.}\
  \bibnamefont {Chiarotti}}, \bibinfo {author} {\bibfnamefont {M.}~\bibnamefont
  {Cococcioni}}, \bibinfo {author} {\bibfnamefont {I.}~\bibnamefont {Dabo}},
  \bibinfo {author} {\bibfnamefont {A.~D.}\ \bibnamefont {Corso}}, \bibinfo
  {author} {\bibfnamefont {S.}~\bibnamefont {de~Gironcoli}}, \bibinfo {author}
  {\bibfnamefont {S.}~\bibnamefont {Fabris}}, \bibinfo {author} {\bibfnamefont
  {G.}~\bibnamefont {Fratesi}}, \bibinfo {author} {\bibfnamefont
  {R.}~\bibnamefont {Gebauer}}, \bibinfo {author} {\bibfnamefont
  {U.}~\bibnamefont {Gerstmann}}, \bibinfo {author} {\bibfnamefont
  {C.}~\bibnamefont {Gougoussis}}, \bibinfo {author} {\bibfnamefont
  {A.}~\bibnamefont {Kokalj}}, \bibinfo {author} {\bibfnamefont
  {M.}~\bibnamefont {Lazzeri}}, \bibinfo {author} {\bibfnamefont
  {L.}~\bibnamefont {Martin-Samos}}, \bibinfo {author} {\bibfnamefont
  {N.}~\bibnamefont {Marzari}}, \bibinfo {author} {\bibfnamefont
  {F.}~\bibnamefont {Mauri}}, \bibinfo {author} {\bibfnamefont
  {R.}~\bibnamefont {Mazzarello}}, \bibinfo {author} {\bibfnamefont
  {S.}~\bibnamefont {Paolini}}, \bibinfo {author} {\bibfnamefont
  {A.}~\bibnamefont {Pasquarello}}, \bibinfo {author} {\bibfnamefont
  {L.}~\bibnamefont {Paulatto}}, \bibinfo {author} {\bibfnamefont
  {C.}~\bibnamefont {Sbraccia}}, \bibinfo {author} {\bibfnamefont
  {S.}~\bibnamefont {Scandolo}}, \bibinfo {author} {\bibfnamefont
  {G.}~\bibnamefont {Sclauzero}}, \bibinfo {author} {\bibfnamefont {A.~P.}\
  \bibnamefont {Seitsonen}}, \bibinfo {author} {\bibfnamefont {A.}~\bibnamefont
  {Smogunov}}, \bibinfo {author} {\bibfnamefont {P.}~\bibnamefont {Umari}}, \
  and\ \bibinfo {author} {\bibfnamefont {R.~M.}\ \bibnamefont {Wentzcovitch}},\
  }\bibfield  {title} {\enquote {\bibinfo {title} {{QUANTUM ESPRESSO: a modular
  and open-source software project for quantum simulations of materials}},}\
  }\href {http://stacks.iop.org/0953-8984/21/i=39/a=395502} {\bibfield
  {journal} {\bibinfo  {journal} {Journal of Physics: Condensed Matter}\
  }\textbf {\bibinfo {volume} {21}},\ \bibinfo {pages} {395502} (\bibinfo
  {year} {2009})}\BibitemShut {NoStop}%
\bibitem [{\citenamefont {Divya}, \citenamefont {Prasad},\ and\ \citenamefont
  {Deepak}(2017)}]{48}%
  \BibitemOpen
  \bibfield  {author} {\bibinfo {author} {\bibnamefont {Divya}}, \bibinfo
  {author} {\bibfnamefont {R.}~\bibnamefont {Prasad}}, \ and\ \bibinfo {author}
  {\bibnamefont {Deepak}},\ }\bibfield  {title} {\enquote {\bibinfo {title}
  {{Recurring polyhedral motifs in the amorphous indium gallium zinc oxide
  network}},}\ }\href {\doibase 10.1002/pssa.201600471} {\bibfield  {journal}
  {\bibinfo  {journal} {physica status solidi (a)}\ }\textbf {\bibinfo {volume}
  {214}},\ \bibinfo {pages} {1600471} (\bibinfo {year} {2017})}\BibitemShut
  {NoStop}%
\bibitem [{\citenamefont {{Noh}}\ \emph {et~al.}(2011)\citenamefont {{Noh}},
  \citenamefont {{Chang}}, \citenamefont {{Ryu}},\ and\ \citenamefont
  {{Lee}}}]{49}%
  \BibitemOpen
  \bibfield  {author} {\bibinfo {author} {\bibfnamefont {H.-K.}\ \bibnamefont
  {{Noh}}}, \bibinfo {author} {\bibfnamefont {K.~J.}\ \bibnamefont {{Chang}}},
  \bibinfo {author} {\bibfnamefont {B.}~\bibnamefont {{Ryu}}}, \ and\ \bibinfo
  {author} {\bibfnamefont {W.-J.}\ \bibnamefont {{Lee}}},\ }\bibfield  {title}
  {\enquote {\bibinfo {title} {{Electronic structure of oxygen-vacancy defects
  in amorphous In-Ga-Zn-O semiconductors}},}\ }\href {\doibase
  10.1103/PhysRevB.84.115205} {\bibfield  {journal} {\bibinfo  {journal} {Phys.
  Rev. B}\ }\textbf {\bibinfo {volume} {84}},\ \bibinfo {eid} {115205}
  (\bibinfo {year} {2011})}\BibitemShut {NoStop}%
\bibitem [{\citenamefont {Cho}\ \emph {et~al.}(2009)\citenamefont {Cho},
  \citenamefont {Song}, \citenamefont {Na}, \citenamefont {Hwang},
  \citenamefont {Jeong}, \citenamefont {Jeong},\ and\ \citenamefont {Mo}}]{50}%
  \BibitemOpen
  \bibfield  {author} {\bibinfo {author} {\bibfnamefont {D.-Y.}\ \bibnamefont
  {Cho}}, \bibinfo {author} {\bibfnamefont {J.}~\bibnamefont {Song}}, \bibinfo
  {author} {\bibfnamefont {K.~D.}\ \bibnamefont {Na}}, \bibinfo {author}
  {\bibfnamefont {C.~S.}\ \bibnamefont {Hwang}}, \bibinfo {author}
  {\bibfnamefont {J.~H.}\ \bibnamefont {Jeong}}, \bibinfo {author}
  {\bibfnamefont {J.~K.}\ \bibnamefont {Jeong}}, \ and\ \bibinfo {author}
  {\bibfnamefont {Y.-G.}\ \bibnamefont {Mo}},\ }\bibfield  {title} {\enquote
  {\bibinfo {title} {{Local structure and conduction mechanism in amorphous
  In–Ga–Zn–O films}},}\ }\href {\doibase 10.1063/1.3103323} {\bibfield
  {journal} {\bibinfo  {journal} {Applied Physics Letters}\ }\textbf {\bibinfo
  {volume} {94}},\ \bibinfo {pages} {112112} (\bibinfo {year}
  {2009})}\BibitemShut {NoStop}%
\bibitem [{\citenamefont {Nomura}\ \emph
  {et~al.}(2007{\natexlab{b}})\citenamefont {Nomura}, \citenamefont {Kamiya},
  \citenamefont {Ohta}, \citenamefont {Uruga}, \citenamefont {Hirano},\ and\
  \citenamefont {Hosono}}]{51}%
  \BibitemOpen
  \bibfield  {author} {\bibinfo {author} {\bibfnamefont {K.}~\bibnamefont
  {Nomura}}, \bibinfo {author} {\bibfnamefont {T.}~\bibnamefont {Kamiya}},
  \bibinfo {author} {\bibfnamefont {H.}~\bibnamefont {Ohta}}, \bibinfo {author}
  {\bibfnamefont {T.}~\bibnamefont {Uruga}}, \bibinfo {author} {\bibfnamefont
  {M.}~\bibnamefont {Hirano}}, \ and\ \bibinfo {author} {\bibfnamefont
  {H.}~\bibnamefont {Hosono}},\ }\bibfield  {title} {\enquote {\bibinfo {title}
  {{Local coordination structure and electronic structure of the large electron
  mobility amorphous oxide semiconductor In-Ga-Zn-O: Experiment and ab initio
  calculations}},}\ }\href {\doibase 10.1103/PhysRevB.75.035212} {\bibfield
  {journal} {\bibinfo  {journal} {Phys. Rev. B}\ }\textbf {\bibinfo {volume}
  {75}},\ \bibinfo {pages} {035212} (\bibinfo {year}
  {2007}{\natexlab{b}})}\BibitemShut {NoStop}%
\bibitem [{\citenamefont {de~Jamblinne~de Meux}\ \emph
  {et~al.}(2015)\citenamefont {de~Jamblinne~de Meux}, \citenamefont {Pourtois},
  \citenamefont {Genoe},\ and\ \citenamefont {Heremans}}]{52}%
  \BibitemOpen
  \bibfield  {author} {\bibinfo {author} {\bibfnamefont {A.}~\bibnamefont
  {de~Jamblinne~de Meux}}, \bibinfo {author} {\bibfnamefont {G.}~\bibnamefont
  {Pourtois}}, \bibinfo {author} {\bibfnamefont {J.}~\bibnamefont {Genoe}}, \
  and\ \bibinfo {author} {\bibfnamefont {P.}~\bibnamefont {Heremans}},\
  }\bibfield  {title} {\enquote {\bibinfo {title} {{Comparison of the
  electronic structure of amorphous versus crystalline indium gallium zinc
  oxide semiconductor: structure, tail states and strain effects}},}\ }\href
  {http://stacks.iop.org/0022-3727/48/i=43/a=435104} {\bibfield  {journal}
  {\bibinfo  {journal} {Journal of Physics D: Applied Physics}\ }\textbf
  {\bibinfo {volume} {48}},\ \bibinfo {pages} {435104} (\bibinfo {year}
  {2015})}\BibitemShut {NoStop}%
\bibitem [{\citenamefont {Orita}\ \emph {et~al.}(2001)\citenamefont {Orita},
  \citenamefont {Ohta}, \citenamefont {Hirano}, \citenamefont {Narushima},\
  and\ \citenamefont {Hosono}}]{53}%
  \BibitemOpen
  \bibfield  {author} {\bibinfo {author} {\bibfnamefont {M.}~\bibnamefont
  {Orita}}, \bibinfo {author} {\bibfnamefont {H.}~\bibnamefont {Ohta}},
  \bibinfo {author} {\bibfnamefont {M.}~\bibnamefont {Hirano}}, \bibinfo
  {author} {\bibfnamefont {S.}~\bibnamefont {Narushima}}, \ and\ \bibinfo
  {author} {\bibfnamefont {H.}~\bibnamefont {Hosono}},\ }\bibfield  {title}
  {\enquote {\bibinfo {title} {{Amorphous transparent conductive oxide
  InGaO$_3$(ZnO)$_m$ (m $\leq$ 4): a Zn4s conductor}},}\ }\href {\doibase
  10.1080/13642810110045923} {\bibfield  {journal} {\bibinfo  {journal}
  {Philosophical Magazine B}\ }\textbf {\bibinfo {volume} {81}},\ \bibinfo
  {pages} {501--515} (\bibinfo {year} {2001})}\BibitemShut {NoStop}%
\bibitem [{\citenamefont {Sanchez-Portal}, \citenamefont {Artacho},\ and\
  \citenamefont {Soler}(1995)}]{58}%
  \BibitemOpen
  \bibfield  {author} {\bibinfo {author} {\bibfnamefont {D.}~\bibnamefont
  {Sanchez-Portal}}, \bibinfo {author} {\bibfnamefont {E.}~\bibnamefont
  {Artacho}}, \ and\ \bibinfo {author} {\bibfnamefont {J.~M.}\ \bibnamefont
  {Soler}},\ }\bibfield  {title} {\enquote {\bibinfo {title} {{Projection of
  plane-wave calculations into atomic orbitals}},}\ }\href {\doibase
  https://doi.org/10.1016/0038-1098(95)00341-X} {\bibfield  {journal} {\bibinfo
   {journal} {Solid State Communications}\ }\textbf {\bibinfo {volume} {95}},\
  \bibinfo {pages} {685 -- 690} (\bibinfo {year} {1995})}\BibitemShut {NoStop}%
\bibitem [{\citenamefont {Kamiya}\ and\ \citenamefont {Hosono}(2010)}]{60}%
  \BibitemOpen
  \bibfield  {author} {\bibinfo {author} {\bibfnamefont {T.}~\bibnamefont
  {Kamiya}}\ and\ \bibinfo {author} {\bibfnamefont {H.}~\bibnamefont
  {Hosono}},\ }\bibfield  {title} {\enquote {\bibinfo {title} {{Material
  characteristics and applications of transparent amorphous oxide
  semiconductors}},}\ }\href {\doibase 10.1038/asiamat.2010.5} {\bibfield
  {journal} {\bibinfo  {journal} {Npg Asia Materials}\ }\textbf {\bibinfo
  {volume} {2}},\ \bibinfo {pages} {15--22} (\bibinfo {year}
  {2010})}\BibitemShut {NoStop}%
\bibitem [{\citenamefont {Hosono}(2006)}]{61}%
  \BibitemOpen
  \bibfield  {author} {\bibinfo {author} {\bibfnamefont {H.}~\bibnamefont
  {Hosono}},\ }\bibfield  {title} {\enquote {\bibinfo {title} {Ionic amorphous
  oxide semiconductors: Material design, carrier transport, and device
  application},}\ }\href {\doibase
  https://doi.org/10.1016/j.jnoncrysol.2006.01.073} {\bibfield  {journal}
  {\bibinfo  {journal} {Journal of Non-Crystalline Solids}\ }\textbf {\bibinfo
  {volume} {352}},\ \bibinfo {pages} {851 -- 858} (\bibinfo {year} {2006})},\
  \bibinfo {note} {amorphous and Nanocrystalline Semiconductors - Science and
  Technology}\BibitemShut {NoStop}%
\bibitem [{\citenamefont {Mulliken}\ \emph {et~al.}(1949)\citenamefont
  {Mulliken}, \citenamefont {Rieke}, \citenamefont {Orloff},\ and\
  \citenamefont {Orloff}}]{54}%
  \BibitemOpen
  \bibfield  {author} {\bibinfo {author} {\bibfnamefont {R.~S.}\ \bibnamefont
  {Mulliken}}, \bibinfo {author} {\bibfnamefont {C.~A.}\ \bibnamefont {Rieke}},
  \bibinfo {author} {\bibfnamefont {D.}~\bibnamefont {Orloff}}, \ and\ \bibinfo
  {author} {\bibfnamefont {H.}~\bibnamefont {Orloff}},\ }\bibfield  {title}
  {\enquote {\bibinfo {title} {{Formulas and Numerical Tables for Overlap
  Integrals}},}\ }\href {\doibase 10.1063/1.1747150} {\bibfield  {journal}
  {\bibinfo  {journal} {The Journal of Chemical Physics}\ }\textbf {\bibinfo
  {volume} {17}},\ \bibinfo {pages} {1248--1267} (\bibinfo {year}
  {1949})}\BibitemShut {NoStop}%
\bibitem [{\citenamefont {Inc.}()}]{55}%
  \BibitemOpen
  \bibfield  {author} {\bibinfo {author} {\bibfnamefont {W.~R.}\ \bibnamefont
  {Inc.}},\ }\href@noop {} {\enquote {\bibinfo {title} {{Mathematica, {V}ersion
  11.0}},}\ }\bibinfo {note} {Champaign, IL, 2016}\BibitemShut {NoStop}%
\bibitem [{59()}]{59}%
  \BibitemOpen
  \href@noop {} {}\bibinfo {note} {The calculated effective mass of {\it
  a}-ZnON is about 25{\%} smaller than effective mass of {\it a}-IGZO, while
  the total orbital overlap integral is 2-3 times higher for {\it a}-ZnON
  compared to {\it a}-IGZO. Both of these do not adequately explain upto an
  order of magnitude higher mobilities observed in {\it a}-ZnON compared to
  {\it a}-IGZO. However, the difference in total orbital overlap integral is
  much closer to observed behavior. We speculate that the orbital overlap
  integral calculations inherently capture some of the physical aspects (such
  as carrier scattering) of carrier transport and hence able to explain the
  observed behavior to a larger extent. On the other hand, effective mass
  calculations are based on DFT calculations, which is applicable to periodic
  structures and does not describe the scattering behavior at all and it may
  miss some of the physical aspects of amorphous structures which are captured
  in total orbital overlap integrals. This may be a subject of further studies
  to describe carrier transport in multi-component amorphous
  semiconductors.}\BibitemShut {Stop}%
\end{thebibliography}%

\end{document}